\PassOptionsToPackage{pdfpagelabels=false}{hyperref}
\documentclass[useAMS,fleqn,usenatbib]{mnras}
\pdfoutput=1
\usepackage{graphicx}
\usepackage{amsmath,amssymb,amstext,color}
\usepackage[T1]{fontenc}
\usepackage{ae,aecompl}
\usepackage[utf8]{inputenc}
\usepackage{newtxtext,newtxmath}
\usepackage[figure,figure*,table,table*]{hypcap}
\usepackage[dvipsnames]{xcolor}
\usepackage{xparse}
\usepackage[title]{appendix}
\usepackage{chngcntr}
\usepackage{multirow}

\newcommand{\aref}[1]{\hyperref[#1]{Appendix~\ref*{#1}}}

\DeclareRobustCommand{\VAN}[3]{#2}
\let\VANthebibliography\thebibliography
\def\thebibliography{\DeclareRobustCommand{\VAN}[3]{##3}\VANthebibliography}

\input{macros.sty}
\input{hyperlink-year-only-natbib-patch}

\title[Count Constraints on Assembly Bias]{Evidence of Galaxy Assembly Bias in SDSS DR7 Galaxy Samples from Count Statistics}

\author[K.~Wang et al.]
{Kuan~Wang$^{1-4}$\thanks{E-mail: \email{kuanwang@umich.edu}},
Yao-Yuan~Mao$^{5}$\thanks{NASA Einstein Fellow; E-mail: \email{yymao.astro@gmail.com}},
Andrew~R.~Zentner$^{3,4}$,
Hong~Guo$^{6}$,
Johannes~U.~Lange$^{7,8}$, \newauthor
Frank~C.~van den Bosch$^{9}$, and
Lorena Mezini$^{3,4}$
\vspace*{6pt} \\ 
$^{1}$Department of Physics, University of Michigan, Ann Arbor, MI 48109, USA\\
$^{2}$Leinweber Center for Theoretical Physics, University of Michigan, Ann Arbor, MI 48109, USA\\
$^{3}$Department of Physics and Astronomy, University of Pittsburgh, Pittsburgh, PA 15260, USA\\
$^{4}$Pittsburgh Particle Physics, Astrophysics, and Cosmology Center (PITT PACC), University of Pittsburgh, Pittsburgh, PA 15260, USA\\
$^{5}$Department of Physics and Astronomy, Rutgers, The State University of New Jersey, Piscataway, NJ 08854, USA\\
$^{6}$Shanghai Astronomical Observatory, CAS, Shanghai 200030, China\\
$^{7}$Department of Astronomy and Astrophysics, University of California, Santa Cruz, CA 95064, USA\\
$^{8}$Kavli Institute for Particle Astrophysics and Cosmology and Department of Physics, Stanford University, CA 94305, USA\\
$^{9}$Department of Astronomy, Yale University, P.O. Box 208101, New Haven, CT 06511, USA\\
\vspace*{-15pt}
}

\date{11 April 2022}
\pubyear{2022}

\begin{document}
\label{firstpage}
\pagerange{\pageref{firstpage}--\pageref{lastpage}}
\maketitle

\begin{abstract}
We present observational constraints on the galaxy--halo connection, focusing particularly on galaxy assembly bias, from a novel combination of counts-in-cylinders statistics, $\Pncic$, with the standard measurements of the projected two-point correlation function, $\wprp$, and number density, $\ngal$, of galaxies.
We measure $\ngal$, $\wprp$ and $\Pncic$ for volume-limited, luminosity-threshold samples of galaxies selected from SDSS DR7, and use them to constrain halo occupation distribution (HOD) models, including a model in which galaxy occupation depends upon a secondary halo property, namely halo concentration. 
We detect significant positive central assembly bias for the $M_r<-20.0$ and $M_r<-19.5$ samples.
Central galaxies preferentially reside within haloes of high concentration at fixed mass. 
Positive central assembly bias is also favoured in the $M_r<-20.5$ and $M_r<-19.0$ samples.
We find no evidence of central assembly bias in the $M_r<-21.0$ sample.
We observe only a marginal preference for negative satellite assembly bias in the $M_r<-20.0$ and $M_r<-19.0$ samples, and non-zero satellite assembly bias is not indicated in other samples.
Our findings underscore the necessity of accounting for galaxy assembly bias when interpreting galaxy survey data, and demonstrate the potential of count statistics in extracting information from the spatial distribution of galaxies, which could be applied to both galaxy--halo connection studies and cosmological analyses.

\end{abstract}



\section{Introduction}

Galaxies form in dark matter haloes, and trace the underlying matter field of the Universe \citep{whiterees78,blumenthal_etal84}. 
In the ongoing campaigns to constrain cosmology and understand galaxy evolution, the spatial distribution of galaxies serves as a rich source of observational information \citep[e.g.,][]{BOSS2020cosmology,kids2021cosmology,DES_Y3_cosmo_clustering_lensing_2021}.
The comparison of cosmological theory against galaxy survey data is dependent on the modelling of galaxy formation and evolution in the large-scale structure \citep[e.g.,][]{desjacques_2018}, specifically, the connection between galaxies and dark matter haloes \citep[see][for a review]{wechsler_tinker18}.
In addition to its use in cosmological analyses, 
the galaxy--halo connection is interesting in its own right, as dark matter haloes dominate the gravitational potential wells within which galaxies 
form and evolve, and halo evolution may regulate many physical processes in galaxies and clusters.

One of the most frequently adopted simplifying assumptions for modelling the galaxy--halo connection is that galaxy properties depend solely on halo mass \citep{mo_etal04,kauffmann_etal04,blanton06}. 
Models based upon this assumption successfully capture the gross behaviour of the large-scale distribution of galaxies.
The success of the ``halo mass-only'' assumption is evident in the halo occupation distribution \citep[HOD;][]{zheng07}, a widely-applied modelling technique which treats the number of central and satellite galaxies in a halo above a luminosity threshold as simple functions of the host halo mass. 
The HOD model is ubiquitous in cosmological analyses due, in part, to both its simplicity and its flexibility. 
However, as increasing amounts of data demand ever greater accuracy in theoretical models of nonlinear clustering, the halo mass-only approximation must submit to increased scrutiny. 

Simulations have long suggested that halo properties other than halo mass can impact galaxy properties.
For example, the assembly histories of haloes tend to correlate with the properties of the galaxies that they host, even at a fixed halo mass. 
Both hydrodynamical galaxy formation simulations \citep[e.g.,][]{Artale_2018,Bose_2019,xu_zheng2020} and semi-analytical models \citep[e.g.,][]{croton_etal07,contreras2019,zehavi2018} show that central galaxy luminosity is higher in older halos at a fixed halo mass.
In addition, both gravity-only \citep[e.g.,][]{zentner_etal05,Mao2015} and hydrodynamical \citep[e.g.,][]{zehavi2018} simulations show that the number of satellite galaxies (or subhaloes in the gravity-only case) is lower in older halos at a fixed halo mass.
These effects are sometimes referred to as \textit{``galaxy assembly bias''} \citep{zentner_etal14}. 

In the presence of different strengths of galaxy assembly bias, the same galaxy clustering signal can, in fact, be interpreted as different halo clustering signals and hence different cosmologies.
As such, analyses of galaxy clustering signals with halo mass-only models, such as the HOD, could be systematically biased \citep[e.g.,][]{wu08,zentner_etal14,McCarthy_2018}, and it is imperative that we include galaxy assembly bias effects properly in galaxy--halo connection models that we use to interpret galaxy survey data. 
Some of the more recent empirical galaxy--halo connection models have attempted to include a tunable strength of galaxy assembly bias \citep[e.g.,][]{Lehmann2017}. 
One notable model is the decorated halo occupation distribution \citep[dHOD;][]{hearin_etal16}, which is based on the HOD formalism, but enables the number of central and satellite galaxies to depend on one additional secondary halo property beyond host halo mass, in our case the halo concentration.
The choice of this additional halo property has also been discussed by several authors, for example, \citet{Xu2021}.

Nevertheless, even equipped with models that include the effects of galaxy assembly bias, it is not straightforward to obtain robust constraints on the strength of galaxy assembly bias in observational data sets. 
In recent years, increasing effort has gone into detecting galaxy assembly bias observationally. 
These efforts are somewhat diverse and involve the application of distinct analysis techniques to a variety of data sets. 
One approach is to construct several different observed galaxy samples at approximately fixed halo mass and compare the large-scale clustering biases among these samples. 
Because it is extremely difficult to build observed samples of galaxies at fixed halo mass (or samples with identical halo mass distributions), results from these approaches have been rather mixed 
\citep[e.g.,][]{tinker_wetzel11,lin_mandelbaum_etal15,miyatake_etal16,more_etal16,dvornik2017,niemiec2018,Obuljen2020,Lin2022}. 
Another approach is to compare observed and simulated galaxy statistics directly (i.e., to adopt a \textit{forward modelling} approach). 
This strategy requires some method for associating galaxies with dark matter haloes, such as a galaxy--halo connection model that includes galaxy assembly bias, e.g., the dHOD. 
In this case, both the choice of the model and the degeneracy between galaxy assembly bias and other model parameters can significantly weaken conclusions, leading to mixed results as well \citep[e.g.,][]{abbas_sheth_06,cooper2010,kauffmann_etal13,wang_etal13,Lehmann2017,vakili_2019,zentner_etal19,salcedo2020,yuan2021}. 
These studies leave open the debate over the existence and potential importance of galaxy assembly bias in studies of galaxy clustering, galaxy evolution, and cosmology.

One way to improve constraints in the forward modelling approach is to include additional galaxy clustering statistics to help break degeneracies and more robustly constrain the galaxy--halo connection. 
In addition to the projected two-point correlation function, $\wprp$, which is commonly used in canonical analyses, other summary statistics have been introduced.
These include the weak lensing signal \citep{cacciato_etal13,lin_mandelbaum_etal15,Lange.etal.19c}, redshift space distortions \citep{reid_etal14, Guo2015wp,Lange.etal.22}, the void probability function \citep[VPF;][]{walsh2019}, and nearest neighbour distributions \citep[kNN-CDF;][]{Banerjee2021_kNNCDF}, each of which provides distinct information. 
In \citet[][hereafter W19]{wang2019_CIC}, we compared several summary statistics, examining the constraining power of different statistic combinations on the galaxy–halo connection.
We forecast that counts-in-cells statistics are superior to the VPF and the weak lensing signal in complementing $\ngal$ and $\wprp$.
We showed that the combination of counts-in-cells statistics with $\ngal$ and $\wprp$ breaks degeneracies, yielding tighter constraints on all HOD parameters, and has the potential to definitively detect or reject assembly bias in existing data.
Similarly, \citet{sinha_etal18} and \citet{szewciw2021} studied a variety of summary statistics and their constraints on the galaxy–halo connection, and found that group multiplicity measures, as well as counts-in-cells statistics, are able to reveal incompleteness in the modelling, and add significantly to the constraining power. 
We aim to build off of these analyses.

In this paper, we select galaxy samples based on luminosity and redshift, from the 7th Data Release of the Sloan Digital Sky Survey \citep[SDSS DR7,][]{DR7_09}, and make independent measurements of our optimal set of statistics, number density $\ngal$, projected two-point function $\wprp$, and counts-in-cylinders $\Pncic$.
This observable set has several virtues besides the improvement of constraining power that we demonstrated in W19.
All three statistics have simple configurations, and are easy to measure from data, without requiring information beyond galaxy positions and luminosities.
The required modelling of these statistics in simulations is likewise minimal.
Both $\wprp$ and $\Pncic$ are projected along the line of sight, and therefore are not particularly sensitive to redshift measurements or models for the peculiar velocities of galaxies (W19). 
As part of our analysis, we conduct careful tests of the impact of observational effects on our statistics, and verify that our model predictions are unbiased with respect to data measurements.
These tests provide a systematic evaluation of the reconstruction procedure of these statistics from a gravity-only $N$-body simulation, which can be generalised to other statistics and other survey data.
We use our measurements to constrain both a mass-only galaxy–halo connection model, the HOD, and an extension of the HOD that incorporates adjustable galaxy assembly bias, the dHOD.

The primary goal of this work is to study galaxy assembly bias in the observed samples using the added information from count statistics.
We find that the inclusion of counts-in-cylinders alongside clustering measurements provides significant evidence for galaxy assembly bias.
Comparing between the clustering-only analysis and the joint analysis of clustering with counts-in-cylinders, the latter also reveals the potential incompleteness in the standard halo mass-only HOD model, and underlines the necessity of properly modelling galaxy assembly bias.
As a subsidiary result, we also produce tighter constraints on the the full parameter space of the galaxy–halo connection, which may aid in the interpretation of simulation results and/or forthcoming survey data.

This paper is organised as follows.
The observational data and cosmological simulation are described in \autoref{sec:datasim}.
In \autoref{sec:measure}, we describe our set of observable statistics, and present measurements from data.
We detail our Markov Chain Monte Carlo fitting in \autoref{sec:mcmc}.
We present our constraints on the galaxy–halo connection, especially the galaxy assembly bias component, in \autoref{sec:results}.
The results are interpreted in \autoref{sec:interpretation} and discussed in \autoref{sec:discussion}.
Conclusions are drawn in \autoref{sec:conclusion}.
In the appendices, we provide technical details and supplementary results. 
In \autoref{sec:validation}, we outline the process for validating the algorithms with which we measure our statistics. 
We include supplementary results for alternative data samples in \autoref{sec:alt_samples}.

\section{Data and Simulation}
\label{sec:datasim}

In this section, we describe the observational data and cosmological simulation that are used in this work.

\subsection{Data and Sample Selection}
\label{sec:data}

In this work, we use galaxy data from the Sloan Digital Sky Survey Data Release 7 \citep[SDSS DR7,][]{DR7_09}.
In particular, we select our samples from the \texttt{bright0} catalogue\footnote{\url{http://sdss.physics.nyu.edu/lss/dr72/bright/0/}}, with $r$-band apparent magnitudes $10.0<m_r<17.6$, in the NYU Value-Added Galaxy Catalog \citep[NYU VAGC,][]{VAGC_05}.
The sample contains galaxies that fall within the survey window, with the bright star-contaminated areas masked out.
We additionally discard sector areas (intersections of tile regions) with low fractions of galaxies that have spectroscopic redshift measurements, i.e., we require that the sector fraction $f_{\rm sector}\geq 0.8$.
We measure the resulting angular area of the data footprint to be approximately 7461 $\rm{deg}^2$, and the total number of galaxies in the sample is 562620.

Due to the finite size of fibres used in the survey, no two targets on the same plate can be closer than 55", which results in a fraction of targeted galaxies not having a measured redshift (known as the fibre collision effect).
Following \citet{zehavi_etal11}, these galaxies are assigned the redshifts of their nearest neighbours.

\begin{figure}
    \centering
    \includegraphics[scale=0.45]{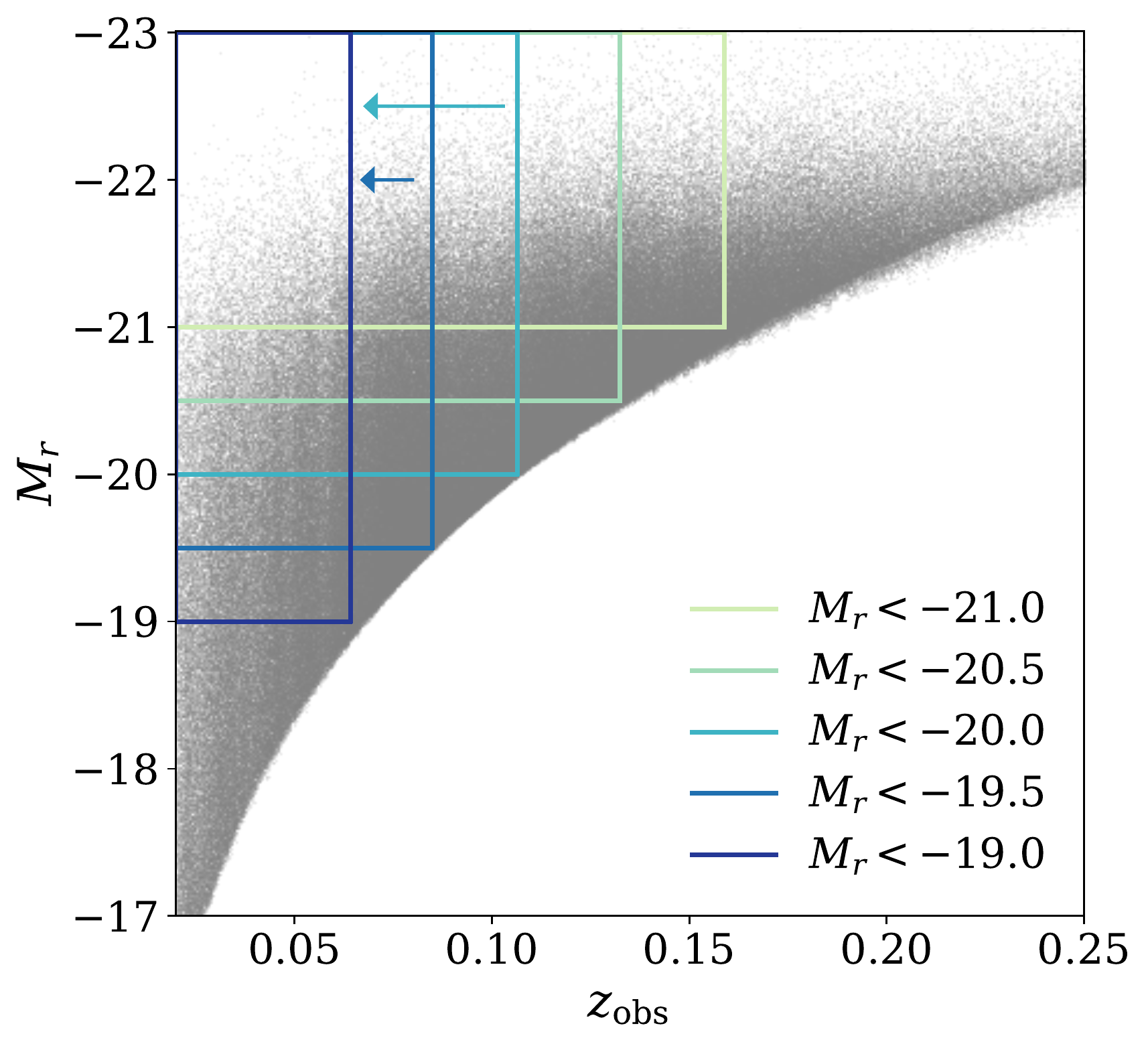}
    \caption{\textbf{Illustration of our volume-limited, luminosity-threshold galaxy samples from the NYU VAGC \texttt{bright0} catalogue.}
    $z_{\rm obs}$ is the observed redshift, where fibre collided galaxies are assigned the redshifts of their nearest neighbours.
    $M_r$ is the {\it r}-band absolute magnitude.
    The scatter points represent the galaxies from the catalogue, and the coloured boxes mark the selection criteria of the different samples, as are labelled in the figure.
    The arrows indicate the alternative $M_r<-20.0$ and $M_r<-19.5$ samples with lower redshift limits that exclude the Sloan Great Wall.}
    \label{fig:sample}
\end{figure}

We select volume-limited, luminosity-threshold samples, based on $r$-band absolute magnitudes that are K-corrected \citep{blanton_roweis07} and passively evolved to the median redshift of the DR7 main galaxy sample, $z=0.1$.
The absolute magnitude values we list as $M_r$ throughout this paper are, in fact, values of $M_r-5\log h$ for $h=1$, which are measured independently of $h$.
We apply a universal lower limit of $z_{\rm min}=0.02$ to all of our samples, and adopt the upper bounds of redshift for each luminosity threshold in \citet{zehavi_etal11}.
In addition to the main samples which maximise the redshift range for each luminosity threshold, these include alternative samples for the $M_r<-20.0$ and $M_r<-19.5$ thresholds with reduced redshift upper bounds that exclude the Sloan Great Wall.
Our sample selection is illustrated in \autoref{fig:sample}, and the details are listed in \autoref{tab:sample}.

\begin{table}[]
    \centering
    \caption{\textbf{Volume-limited samples with luminosity threshold.}
    All of the samples have minimum redshift $z_{\rm min}=0.02$.
    Redshift upper bounds are listed in terms of their products with the speed of light, $c$.
    The galaxies in each sample satisfy the conditions $z_{\rm min}\leq z_{\rm obs} < z_{\rm max}$ and $M_r<M_{r, \rm max}$.
    The second part of the table lists alternatively selected samples for the $M_r<-20.0$ and $M_r<-19.5$ thresholds (marked with asterisks), which exclude the cosmic structure known as the Sloan Great Wall by adopting the shallower redshift bounds.}
    \label{tab:sample}
    \begin{tabular}{c|c|c}
    \multicolumn{3}{c}{} \\
		\hline\hline
        $M_{r, \rm max}$ & $cz_{\rm max}[\kms]$ & $N_{\rm gal}$ \\
        \hline
        & Main samples & \\
		\hline
        -21.0 & 47650 & 82263 \\ 
        -20.5 & 39700 & 130707 \\  
        -20.0 & 31900 & 140149 \\
        -19.5 & 25450 & 131322\\ 
        -19.0 & 19250 & 76442\\  
        \hline
        & Alternative samples & \\ 
        \hline
        $-20.0^*$ & 19250 & 29951\\
        $-19.5^*$ & 19250 & 51007\\
		\hline\hline
    \end{tabular}
\end{table}

\subsection{Simulation}
\label{sec:sim}

Our forward modelling analyses are based on the Small MultiDark Planck Simulation (SMDPL), which is a gravitational {\it N}-body simulation that belongs to the series of MultiDark simulations with Planck cosmology \citep{klypin_etal16}. 
The Small MultiDark Planck Simulation has a cubic volume of side length 400 $\Mpch$, which is comparable to the volume of our $M_r<-21.0$ data sample.
The cosmological parameters adopted are $\Omega_{\Lambda}=0.6929$, $\Omega_{\rm m}=1-\Omega_{\Lambda}=0.3071$, $\Omega_{\rm b}=0.0482$, $h=0.6777$, $n_{\rm s}=0.96$, and $\sigma_8=0.8228$. 
We assume this cosmology in our analyses throughout this paper.
The simulation is evolved from $z_{\rm initial}=120$, with $3840^3$ particles, implying a particle mass resolution of $m_{\rm p} = 9.63 \times 10^7 \Msunh$.
We use halo catalogues extracted from the $z=0.1$ snapshot using the \texttt{Rockstar} halo-finder \citep{rockstar}, downloaded from \url{https://www.cosmosim.org}.
We truncate the halo peak mass $M_{\rm peak}$ at $300\times m_{\rm p} = 2.889\times 10^{10}\Msunh$, below which the probability for a halo to host galaxies above our luminosity thresholds is very low ($\lesssim 10^{-4}$).

\section{Measurements}
\label{sec:measure}

In this section, we present measurements of the observable statistics that we use from the SDSS galaxy samples.

\subsection{Observable Statistics}
\label{sec:stat}

We employ summary statistics to extract information from the spatial distribution of galaxies. 
In W19, we showed that the combination of the projected two-point correlation function $\wprp$ and the counts-in-cylinders statistic $\Pncic$ yields tight constraints on the galaxy–halo connection by breaking degeneracies in the model parameter space.
In this work, we elect to measure these statistics along with the galaxy number density $\ngal$.

The projected two-point correlation function is defined by 
\begin{equation}
\label{eq:wp}
\wprp = 2\int_{0}^{\pimax}d\pi \ \xi(\rp,\pi),
\end{equation}
where $\xi(\rp,\pi)$ is the excess probability of finding galaxy pairs with projected and line-of-sight separations $\rp$ and $\pi$, respectively.
Considering the depth of the data samples and the size of the simulation, we choose $\pimax=40\Mpch$, to include most correlated pairs and reduce the impact of peculiar velocities, while excluding distant and less correlated pairs. 
We compute $\wprp$ in 12 logarithmically-spaced, radial bins between $\rp=0.158\Mpch$ and $\rp=39.81\Mpch$.

The counts-in-cylinders statistic is the probability distribution of the number of companions found in cylinders around galaxies.
As was done in W19, we centre a cylinder of transverse radius $\rcic=2\Mpch$ and line-of-sight half-length $L=10\Mpch$ on each galaxy in the sample, and count the number of companion galaxies that fall within the cylinder.
We then estimate a probability distribution of companion number, $\Pncic$, which is the probability that any galaxy has $\Ncic$ companions within the cylinder.
When characterising $\Pncic$, we bin $\Ncic$ values linearly on the lower end and logarithmically on the higher end, as listed in \autoref{tab:datanwc}.

\subsection{Measurement Algorithm}
\label{sec:data_algorithm}

Our algorithm for measuring the observable set is as follows.
The number density, $\ngal$, is calculated by dividing the number of galaxies by the volume within the survey footprint and redshift range of each sample. 
Both $\wprp$ and $\Pncic$ are based on pair counting, and we first convert the angular and redshift separations between galaxies to transverse and line-of-sight separations, according to the cosmological model that we adopt.

For $\wprp$, we use the Landy-Szalay estimator \citep{landyszalay93},
\begin{equation}
\label{eq:xi_ls}
\hat{\xi}_{\rm LS} = \frac{\rm{DD}-2\rm{DR}+\rm{RR}}{\rm{RR}},
\end{equation}
where DD is the normalised galaxy–galaxy pair count, RR is the normalised random–random pair count, and DR is the normalised cross pair count between galaxies and randoms.
Randoms are drawn from a uniform distribution in the redshift range and footprint of the survey, to account for the geometry of the survey volume.
$\xi$ is integrated along the line of sight to $\pimax$.
The calculation is done using the \textsc{Corrfunc} package \citep{corrfunc}.

For $\Pncic$, we impose additional criteria on the galaxies used as cylinder centres, such that the sampling of companions is sufficient in the neighbourhood of each cylinder centre.
We define the angular completeness $f_{\rm AC}$ of a galaxy to be the fraction of the circular area with radius $\rcic=2\Mpch$ around it that falls inside the survey footprint.
For cylinder centres we require that the angular completeness around them to be above 0.9, and cylinders centred on them to be completely within the redshift ranges of the volume-limited samples.
For each of the cylinders that satisfy these requirements, we then count companions that belong to the same volume-limited, luminosity-threshold sample.
We upscale the count numbers by $1/f_{\rm AC}$ to account for any angular incompleteness, which results in non-integer counts.
The counts are binned to yield the probability distribution, $\Pncic$.

For all the measurements, we adopt the nearest-neighbour corrected redshifts for fibre collided galaxies that do not have spectroscopic redshifts.

\subsection{Covariance Estimation}
\label{sec:cov_est}

\begin{figure*}
    \centering
    \includegraphics[scale=0.7]{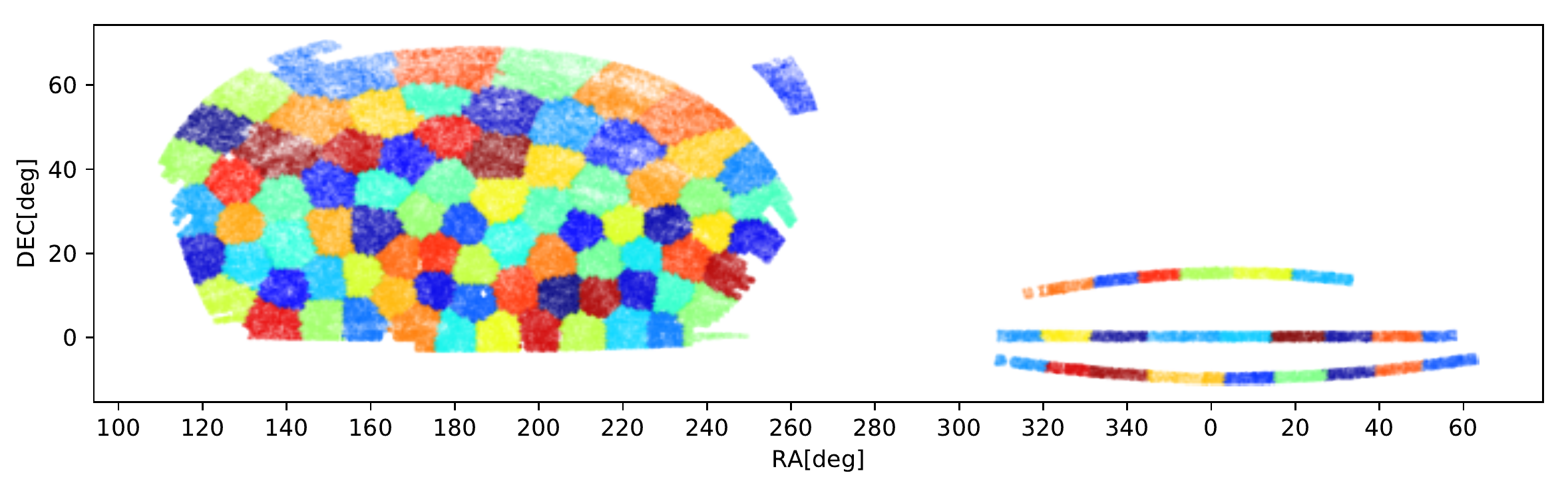}
    \caption{\textbf{SDSS footprint divided into jackknife cells.}
    Objects in different cell are plotted in different colours.
    All the cells are similar in size, except some in the three stripes away from the main footprint, which are smaller than average.}
    \label{fig:jkcells}
\end{figure*}

We calculate the jackknife covariance, which provides an estimate of the uncertainty due to the finite volume of the survey.
We use the iterative method described in \citet{Zhou2021}\footnote{Code available at \url{https://github.com/rongpu/pixel_partition}}, to divide the survey footprint into subareas of similar sizes.
We show a map of the cells in \autoref{fig:jkcells}, where galaxies in different cell are plotted in different colours.
For each sample, we exclude one cell at a time in our measurement to obtain an ensemble of jackknife subsamples.
In particular, for $\ngal$, galaxies are excluded if they fall in the excluded cell, for $\wprp$, we exclude a pair of objects if either or both are in the excluded cell, whereas for $\Pncic$, we keep all the galaxies as companions, and only exclude cylinder centres that fall in the excluded cell from the probability distribution.
We then record the covariance matrix across the ensemble as our jackknife covariance.

\subsection{Measurement Results}
\label{sec:data_stats}

\begin{figure*}
    \centering
    \label{fig:datanw}
    \includegraphics[scale=0.38]{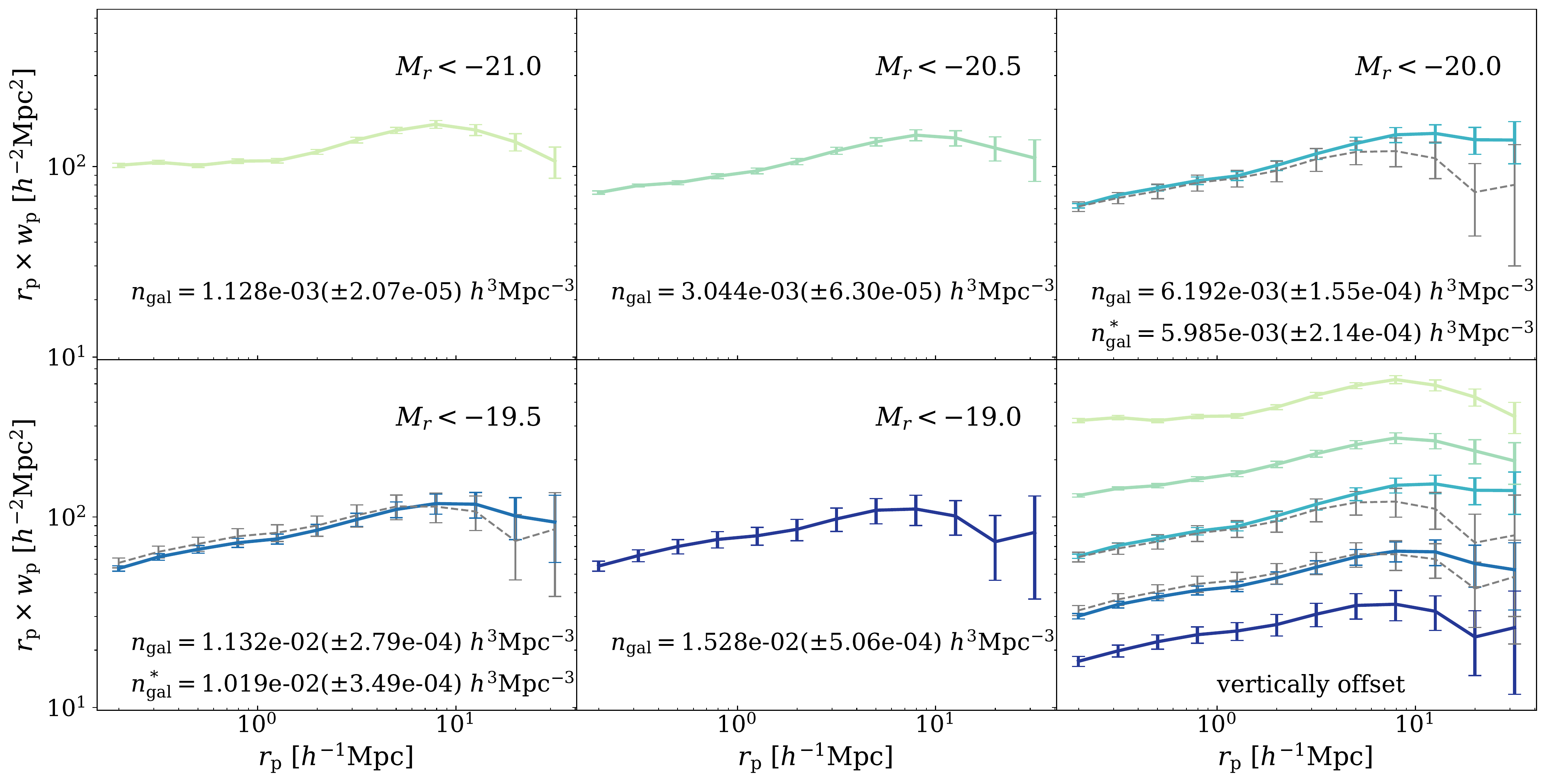}
    \caption{\textbf{Measurement of the galaxy number density $\ngal$ and the projected two-point function $\wp$ from SDSS data.}
    The two point function is shown in the form of $\rp\times\wp$ instead of $\wp$ for legibility.
    Each of the first five panels shows the measurement for one luminosity sample with jackknife error bars that represent the cosmic variance.
    For the $-20.0$ and $-19.5$ thresholds, results for the alternative samples without the Sloan Great Wall are shown in the respective panels as values with asterisks and grey lines.
    The bottom right panel shows $\rp\times\wprp$ of the five thresholds together, with vertical offsets of 0.25 dex each, starting from the $-20.0$ sample, for visual clarity.}
\end{figure*}

\begin{figure*}
    \centering
    \label{fig:datac}
    \includegraphics[scale=0.38]{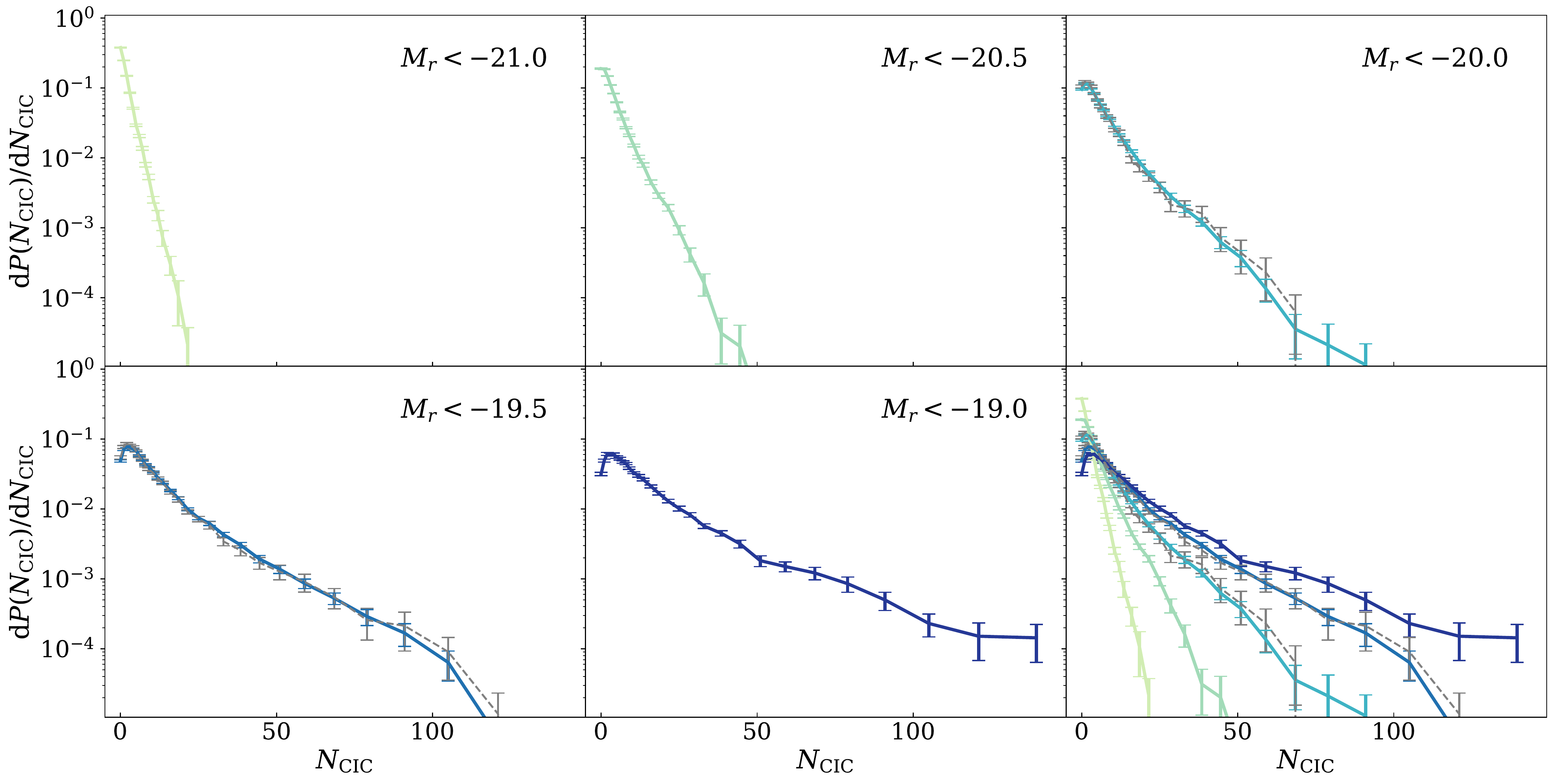}
    \caption{\textbf{Measurement of the counts-in-cylinders statistic $\Pncic$ from SDSS data.}
    Similar to \autoref{fig:datanw}, each of the first five panels shows the measurement for one luminosity sample with jackknife error bars that represent the cosmic variance.
    The statistic is represented as the probability distribution of $\Ncic$, normalised by the bin widths.
    For the $-20.0$ and $-19.5$ thresholds, results for the alternative samples without the Sloan Great Wall are shown in the respective panels as grey lines. 
    The bottom right panel shows the $\Pncic$ of the five thresholds together, with no vertical offset.}
\end{figure*}

The $\wprp$ values we measure from data are shown in \autoref{fig:datanw} in the form of $\rp\times\wp$, where each luminosity threshold is plotted individually in the first five panels, and compared alongside each other in the last panel.
The error bars show the jackknife error of each $\rp$ bin.
For the $M_r<-20.0$ and $M_r<-19.5$ thresholds, the alternative samples with the shallower redshift ranges are plotted in grey.
Our measurements for $\wprp$ are consistent with the measurements in \citet{zehavi_etal11}.
Number densities are also shown in text in the corresponding panels, with jackknife errors included in parentheses.
$\ngal$ values for the alternative samples are marked with asterisks.
The $\Pncic$ statistics are similarly shown in \autoref{fig:datac}.
The values of the statistics are listed in \autoref{tab:datanwc}.

Brighter samples have lower number densities, but generally stronger clustering.
The $\Pncic$ values in all the bins sum up to unity, as $\Pncic$ is a probability distribution.
Brighter galaxies are rarer, and tend to have fewer companions, resulting in higher probabilities of smaller $\Ncic$, and lower probabilities of larger $\Ncic$.
The samples that have smaller volumes have larger jackknife errors for all of the observables.

For the $M_r<-20.0$ and $M_r<-19.5$ luminosity thresholds, the alternative redshift limit that excludes the Sloan Great Wall results in lower number densities and weaker large-scale clustering, as is expected.
For both thresholds, excluding the Great Wall increases the probability of galaxies having only a few companions in cylinders, reduces that of intermediate numbers of companions, but shows an increase in the probability of having very large numbers of companions.
However, neither the increase of $\Pncic$ at the large $\Ncic$ end nor the increase of $\wprp$ are statistically significant.

\begin{table*}[]
    \centering
    \caption{\textbf{Measured values of the observable statistics from SDSS data.}
    $\ngal$, $\wprp$, and $\Pncic$ measurements are listed in three respective sections from top to bottom.
    The number density values $\ngal$ are listed in units of $10^{-3}h^3{\rm Mpc}^{-3}$.
    The leftmost column shows the $\rp$ bin centre in units of $\Mpch$ for $\wp$, and the edges of each bin in $\Ncic$ for $\Pncic$.
    The values listed for $\wprp$ are in units of $\Mpch$, and the values listed for $\Pncic$ are the probability in each bin, which would sum up to unity.
    Jackknife errors are shown in brackets.}
    \label{tab:datanwc}
    \begin{tabular}{|l|lllllll|}
    \hline\hline
    & $M_r<-21.0$ & $M_r<-20.5$ & $M_r<-20.0$ & $M_r<-19.5$ & $M_r<-19.0$ & $M_r<-20.0^*$ & $M_r<-19.5^*$ \\ 
    \hline
    & \multicolumn{7}{l}{\hspace{170pt}$\ngal\left[10^{-3}h^3{\rm Mpc}^{-3}\right]$}\\\hline
    & 1.128(0.021) & 3.044(0.063) & 6.192(0.155) & 11.32(0.279) & 15.28(0.506) & 5.985(0.214) & 10.19(0.349) \\
    \hline
    $\rp\left[\Mpch\right]$ & \multicolumn{7}{l}{\hspace{174pt}$\wprp\left[\Mpch\right]$}\\\hline
    0.200 &  508.36(13.00) &  365.33(7.38) &  311.92(8.25) &  268.89(8.67) &  276.92(16.64) &  308.88(17.98) &  288.03(17.14) \\
    0.316 &  332.75(8.30) &  251.26(4.73) &  224.05(6.71) &  195.09(7.68) &  198.23(14.26) &  216.09(14.45) &  207.49(14.93) \\
    0.501 &  201.96(4.74) &  163.95(4.17) &  154.16(5.82) &  134.95(6.13) &  139.73(12.11) &  148.19(12.91) &  144.09(12.32) \\
    0.794 &  134.25(3.51) &  111.89(3.23) &  106.07(4.84) &  92.18(4.97) &  95.93(9.47) &  103.46(9.92) &  99.67(9.62) \\
    1.259 &  85.17(2.32) &  75.32(2.58) &  70.86(3.78) &  60.96(3.72) &  63.29(6.79) &  68.98(6.96) &  65.68(6.67) \\
    1.995 &  59.77(1.76) &  53.23(2.04) &  50.78(3.06) &  42.72(3.13) &  43.15(5.57) &  47.56(5.88) &  45.20(5.55) \\
    3.162 &  43.54(1.51) &  38.23(1.62) &  36.87(2.46) &  30.64(2.47) &  30.89(4.32) &  34.55(4.74) &  32.29(4.31) \\
    5.012 &  30.86(1.13) &  26.89(1.35) &  26.36(2.04) &  21.87(2.07) &  21.65(3.29) &  23.78(3.38) &  22.65(3.34) \\
    7.943 &  20.96(1.00) &  18.39(1.21) &  18.47(1.66) &  14.80(1.79) &  13.86(2.50) &  15.16(2.61) &  14.27(2.54) \\
    12.59 &  12.35(0.81) &  11.21(1.03) &  11.84(1.32) &  9.27(1.42) &  8.03(1.65) &  8.77(1.91) &  8.48(1.74) \\
    19.95 &  6.75(0.70) &  6.26(0.91) &  6.92(1.11) &  5.07(1.26) &  3.72(1.38) &  3.67(1.51) &  3.75(1.41) \\
    31.62 &  3.37(0.63) &  3.50(0.86) &  4.35(1.09) &  2.97(1.15) &  2.63(1.45) &  2.53(1.58) &  2.73(1.52) \\
    \hline
    $\Ncic$ & \multicolumn{7}{l}{\hspace{187pt}$\Pncic$}\\\hline
    -0.5, 0.5    &  0.379(4.82e-3) &  0.189(3.87e-3) &  0.096(2.82e-3) &  0.049(1.82e-3) &  0.032(1.79e-3) &  0.106(4.95e-3) &  0.054(2.97e-3) \\ 
    0.5, 1.5     &  0.249(2.49e-3) &  0.187(2.80e-3) &  0.117(2.91e-3) &  0.071(2.48e-3) &  0.049(2.53e-3) &  0.123(5.43e-3) &  0.077(3.82e-3) \\ 
    1.5, 2.5     &  0.149(2.07e-3) &  0.148(1.87e-3) &  0.113(2.43e-3) &  0.078(2.47e-3) &  0.061(2.98e-3) &  0.117(4.86e-3) &  0.086(3.51e-3) \\ 
    2.5, 3.5     &  0.085(1.76e-3) &  0.111(1.49e-3) &  0.099(1.96e-3) &  0.077(2.22e-3) &  0.060(2.87e-3) &  0.106(3.92e-3) &  0.081(3.22e-3) \\ 
    3.5, 4.5     &  0.052(1.62e-3) &  0.083(1.27e-3) &  0.084(1.59e-3) &  0.072(2.04e-3) &  0.061(2.50e-3) &  0.084(2.93e-3) &  0.077(3.25e-3) \\ 
    4.5, 5.5     &  0.029(1.21e-3) &  0.063(1.19e-3) &  0.069(1.14e-3) &  0.068(1.75e-3) &  0.055(2.24e-3) &  0.068(2.50e-3) &  0.068(2.47e-3) \\ 
    5.5, 6.5     &  0.021(1.08e-3) &  0.046(1.05e-3) &  0.058(1.05e-3) &  0.058(1.31e-3) &  0.052(2.30e-3) &  0.055(2.87e-3) &  0.056(2.01e-3) \\ 
    6.5, 7.5     &  0.014(8.48e-4) &  0.036(1.09e-3) &  0.049(1.06e-3) &  0.050(1.03e-3) &  0.047(1.76e-3) &  0.048(2.42e-3) &  0.050(1.88e-3) \\ 
    7.5, 8.5     &  0.008(5.99e-4) &  0.027(8.79e-4) &  0.040(9.48e-4) &  0.043(1.05e-3) &  0.045(1.60e-3) &  0.039(2.25e-3) &  0.041(1.61e-3) \\ 
    8.5, 9.5     &  0.005(4.64e-4) &  0.021(8.50e-4) &  0.036(9.49e-4) &  0.039(8.89e-4) &  0.038(1.54e-3) &  0.036(2.36e-3) &  0.037(1.59e-3) \\ 
    9.5, 11.5    &  0.005(5.55e-4) &  0.030(1.38e-3) &  0.055(1.60e-3) &  0.067(1.70e-3) &  0.066(1.98e-3) &  0.050(2.98e-3) &  0.065(2.93e-3) \\ 
    11.5, 12.5   &  0.002(2.56e-4) &  0.010(6.21e-4) &  0.021(7.80e-4) &  0.028(8.75e-4) &  0.030(1.29e-3) &  0.023(2.38e-3) &  0.027(1.41e-3) \\ 
    12.5, 14.5   &  0.001(3.72e-4) &  0.016(1.10e-3) &  0.035(1.36e-3) &  0.048(1.53e-3) &  0.053(2.28e-3) &  0.033(2.71e-3) &  0.047(2.61e-3) \\ 
    14.5, 17.5   &  0.001(2.71e-4) &  0.013(1.05e-3) &  0.037(1.69e-3) &  0.055(2.01e-3) &  0.063(3.12e-3) &  0.028(2.97e-3) &  0.052(3.21e-3) \\ 
    17.5, 19.5   &  0.000(1.35e-4) &  0.006(5.12e-4) &  0.018(1.07e-3) &  0.028(1.28e-3) &  0.034(1.93e-3) &  0.014(1.75e-3) &  0.027(2.26e-3) \\ 
    19.5, 23.5   &  0.000(6.37e-5) &  0.008(8.23e-4) &  0.024(1.83e-3) &  0.040(1.74e-3) &  0.052(3.00e-3) &  0.022(2.99e-3) &  0.036(2.44e-3) \\ 
    23.5, 26.5   &               0 &  0.003(4.13e-4) &  0.012(1.13e-3) &  0.022(1.25e-3) &  0.030(2.39e-3) &  0.012(2.00e-3) &  0.022(1.90e-3) \\ 
    26.5, 30.5   &               0 &  0.002(3.80e-4) &  0.011(1.13e-3) &  0.025(1.69e-3) &  0.033(2.35e-3) &  0.009(1.72e-3) &  0.023(2.80e-3) \\ 
    30.5, 35.5   &               0 &  0.001(2.83e-4) &  0.009(1.09e-3) &  0.021(1.63e-3) &  0.028(2.19e-3) &  0.010(2.49e-3) &  0.017(2.24e-3) \\ 
    35.5, 41.5   &               0 &  0.000(1.19e-4) &  0.007(9.51e-4) &  0.018(1.66e-3) &  0.027(2.45e-3) &  0.010(2.49e-3) &  0.015(2.55e-3) \\ 
    41.5, 47.5   &               0 &  0.000(1.21e-4) &  0.004(7.29e-4) &  0.012(1.38e-3) &  0.019(2.42e-3) &  0.004(1.66e-3) &  0.010(1.92e-3) \\ 
    47.5, 54.5   &               0 &  0.000(1.87e-5) &  0.003(6.83e-4) &  0.010(1.30e-3) &  0.013(2.24e-3) &  0.003(1.55e-3) &  0.009(2.09e-3) \\ 
    54.5, 63.5   &               0 &               0 &  0.001(4.34e-4) &  0.008(1.21e-3) &  0.014(2.20e-3) &  0.002(1.26e-3) &  0.008(2.33e-3) \\ 
    63.5, 73.5   &               0 &               0 &  0.000(2.22e-4) &  0.005(9.92e-4) &  0.012(2.46e-3) &  0.001(4.72e-4) &  0.005(1.77e-3) \\ 
    73.5, 84.5   &               0 &               0 &  0.000(2.32e-4) &  0.003(8.14e-4) &  0.009(2.32e-3) &               0 &  0.003(1.36e-3) \\ 
    84.5, 97.5   &               0 &               0 &  0.000(1.43e-4) &  0.002(7.83e-4) &  0.006(1.91e-3) &          0 &  0.003(1.57e-3) \\ 
    97.5, 112.5  &               0 &               0 &               0 &  0.001(4.39e-4) &  0.003(1.24e-3) &               0 &  0.001(8.22e-4) \\ 
    112.5, 129.5 &               0 &               0 &               0 &  0.000(7.32e-5) &  0.003(1.41e-3) &               0 &  0.000(1.98e-4) \\ 
    129.5, 149.5 &               0 &               0 &               0 &               0 &  0.003(1.59e-3) &               0 &               0 \\  
    \hline\hline
    \end{tabular}
\end{table*}

\section{Fitting Models to Observational Data}
\label{sec:mcmc}

In this section, we describe how we construct our models and the approach for fitting the clustering and counts-cylinders measurements using galaxy–halo connection models with and without galaxy assembly bias components.

\subsection{Galaxy–Halo Connection Models}
\label{sec:gh_model}

We fit two galaxy–halo connection models –– one halo mass-only model, in which the galaxy content of a halo depends only upon the mass of the halo, and an alternative model that incorporates galaxy assembly bias –– to the data measurement.
We choose the halo occupation distribution \citep[HOD;][]{zheng07} model for the former, and the decorated halo occupation distribution \citep[dHOD;][]{hearin_etal16} model for the latter.
We give a brief description of the models in this subsection, and encourage the interested reader to refer back to the original papers, as well as Section 2.2 of W19, for details and motivations for the specific choices we make.

In the standard HOD, the number of galaxies in a given halo is determined solely by the mass of the halo, $\Mvir$.
The dependence of the {\em mean} occupation on halo mass is modulated by 5 parameters. 
The mean number of central and satellite galaxies in halos of fixed mass are modelled separately as
\begin{eqnarray}
\langle \Ncen \vert \Mvir \rangle & = & \frac{1}{2} \left( 1 + 
\mathrm{erf} \left[\frac{\log\Mvir-\log\Mmin}{\slogM}  \right]  \right), \label{eq:nstd1}\\
\langle \Nsat \vert \Mvir  \rangle & = & \left(\frac{\Mvir - \Mzero}{\Mone}\right)^{\alpha} \times \langle \Ncen \vert \Mvir \rangle.
\label{eq:nstd2}
\end{eqnarray}
The number of central galaxies assigned to any particular halo, $\Ncen$, is drawn from a Bernoulli distribution with a mean given by Eq.~(\ref{eq:nstd1}). 
Therefore, any individual halo can have, at most, one central galaxy. 
The number of satellite galaxies assigned to any halo, $\Nsat$, is drawn from a Poisson distribution with mean given by Eq.~(\ref{eq:nstd2}).
While the probability for a halo to host satellite galaxies is modulated by the mean central occupation, a central galaxy is not strictly required for satellites to be present for individual haloes.

The decorated HOD has an additional galaxy assembly bias component.
This is implemented by introducing a dependence of the mean numbers of satellite and central galaxies on a secondary halo property $x$ (any property aside from mass is a secondary property).
At each halo mass, we categorise the haloes into two sub-populations, with a pivot value of $x_{\mathrm{piv}}$,
\begin{eqnarray}
\langle \Ngal \vert \Mvir, x>x_{\mathrm{piv}} \rangle & = & 
\langle \Ngal \vert \Mvir \rangle + \delta \Ngal, \label{eq:ngaldec1}\\
\langle \Ngal \vert \Mvir, x \le x_{\mathrm{piv}} \rangle & = &
\langle \Ngal \vert \Mvir \rangle - \delta \Ngal,
\label{eq:ngaldec2}
\end{eqnarray}
where $\Ngal$ is either $\Ncen$ or $\Nsat$.
In this work, we choose the NFW \citep{nfw97} halo concentration $\cvir$ as the secondary halo property, that is, $x=\cvir$, and choose $x_{\mathrm{piv}}$ to be the median value of $\cvir$ in each mass bin.
The variation is then given by
\begin{eqnarray}
\delta \Ncen & = & \Acen \, \mathrm{min}
\left[\langle \Ncen \vert \Mvir \rangle, 1-\langle \Ncen \vert \Mvir \rangle\right],\label{eq:dngal1}\\
\delta \Nsat & = & \Asat \, \langle \Nsat \vert \Mvir \rangle,
\label{eq:dngal2}
\end{eqnarray}
where the two assembly bias parameters, $\Acen$ and $\Asat$, range from -1 to 1, and the special case of $\Acen=0$ and $\Asat=0$ is equivalent to the standard HOD (i.e., no assembly bias).
Throughout the text, we refer to $\Acen$ as the central assembly bias parameter, and $\Asat$ as the satellite assembly bias parameter.
The existence of central and satellite assembly biases (or the lack thereof) discussed in this work is limited to this specific model choice. 

For the purposes of this study, we assign galaxies positions and velocities relative to their halo centres in the default manner specified within the \texttt{halotools} software.
In particular, central galaxies are placed at the centres of their host haloes and inherit the host halo’s peculiar velocity as identified by the halo finder.
Satellite galaxies are distributed according to a spherically-symmetric NFW profile with the same concentration as the dark matter distribution.
Satellite peculiar velocities are modelled assuming isotropy, and the radial velocity follows a Gaussian distribution, whose first moment is the host halo velocity, and second moment is the solution of the isotropic Jeans equation \citep{klypin99a}, for an NFW profile with the same concentration as the dark matter in the halo.

\subsection{Model Prediction of Statistics}
\label{sec:validation_brief}

We predict $\ngal$, $\wprp$, and $\Pncic$ from the galaxy–halo connection models using the $z=0.1$ snapshot of the cubic SMDPL simulation.
The details of the measurement of these statistics are described in \autoref{sec:cube_algorithm}.
To ensure that the predictions from the cubic simulation volume are consistent with observational estimations of the statistics, we need to understand the observational effects that enter the SDSS data measurements.
Apart from the corrections on absolute magnitudes already made in the NYU VAGC, we consider major observational effects in SDSS (see \autoref{sec:data} for details) including i) the light cone geometry with a non-trivial footprint, ii) the fibre collision effect and other failures to retrieve galaxy redshifts, and iii) redshift space distortion due to peculiar velocities of galaxies.
While iii) can be incorporated into the cubic simulation by modelling mock galaxy velocities, i) and ii) are unique to the geometry of observational data.
We construct cone mocks that incorporate these observational effects, and use them to validate our measurement algorithms, demonstrating that our statistic estimates from the observational data and the cubic simulation are consistent within error.
The details of the cone construction and the validation process are described in \autoref{sec:validation}.

\subsection{Covariance Matrix}
\label{sec:mccov}

\begin{figure*}
    \centering
    \includegraphics[scale=0.34]{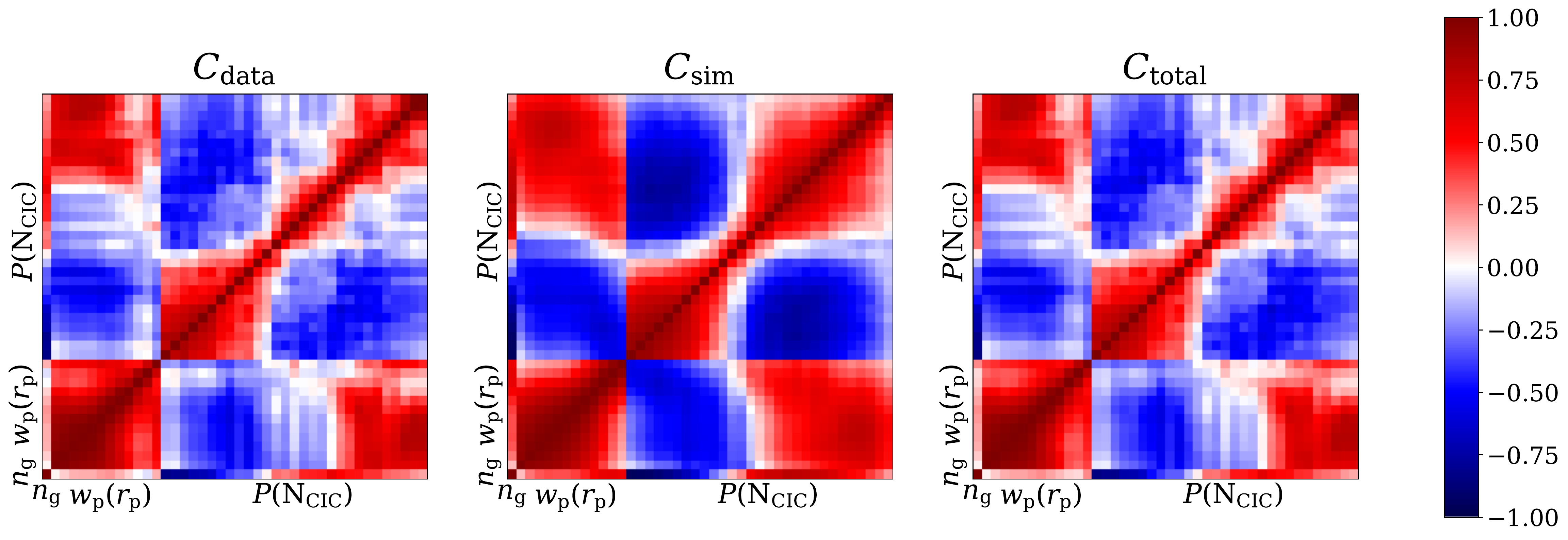}
    \caption[Covariance matrix for Mr<-19.0 sample.]
    {\textbf{Normalised covariance matrices (i.e., correlation matrices) for the $M_r<-19.0$ sample.}
    The left panel is the jackknife covariance of the SDSS data sample, the middle panel that of the SMDPL simulation, and the right panel shows the sum of the two.
    In the figure, the $\rp$ bins and the $\Ncic$ bins both increase from left to right and from bottom to top.
    The diagonal elements of the correlation matrices are 1 by construction.}
    \label{fig:cov19p0}
\end{figure*}

In fitting data measurements with mock measurements from the simulation, we need to account for both the uncertainty in the observation and the uncertainty in the mock estimation.
Therefore, the covariance matrix that we use for the fitting is the sum of the data component and the theory component.
In \autoref{fig:cov19p0}, we show the normalised covariance matrix for the $M_r<-19.0$ sample as an example.
The two panels on the left are the jackknife covariances from the SDSS data and the SMDPL mock separately, and the rightmost panel is the sum of both.
The estimation of the SDSS data covariance is detailed in \autoref{sec:cov_est}, and the estimation of the SMDPL mock covariance is detailed in \autoref{sec:cube_cov}.
In plotting the matrices, they are normalised by the diagonal elements, such that all the diagonal elements are normalised to 1 by construction, and the off-diagonal elements range between -1 and 1, which are colour coded in the figure.
The data covariance is noisier due to its smaller volume than the simulation covariance.

We treat the set of observables as one data vector, and include covariance terms across observables.
For each luminosity sample, in both the data and the mock components of the covariance, $\ngal$ is positively correlated with $\wp$ in general and positively correlated with counts of companions in cylinders.
Within the same galaxy population, galaxy clustering is stronger in regions of space that are slightly denser than average. 
The correlations between $\wprp$ values across different scales are positive within the range we investigate, and their correlations with $\Pncic$ approximately follow those of $\ngal$.
On the other hand, $\Pncic$ values at the higher and lower ends are anti-correlated by construction, as these probabilities sum up to unity.

\subsection{MCMC Fitting}
\label{sec:fitting}

We use the Markov Chain Monte Carlo (MCMC) method to infer parameter constraints from the measurements. 
The covariance matrices we use in the fit include the cross covariances between different statistics, as is described in \autoref{sec:mccov}.
When sampling the parameter space, we populate the haloes in the simulation with galaxies according to each parameter set, make mock measurements on the resulting galaxy catalogue, and compute the corresponding likelihood.
We assume likelihood $\mathcal{L}\propto e^{-\chi^2/2}$, with
\begin{equation}
    \chi^2=\sum_{i,j} \Delta f_i [\mathbf{C}^{-1}]_{ij} \Delta f_j,
\end{equation}
where $f_i$ and $f_j$ are the $i$th and $j$th element of the joint statistic vector $\mathbf{f}$, and $\mathbf{C}$ is the full covariance.
Because the three types of statistics, $\ngal$, $\wprp$, and $\Pncic$, have a wide range of values, we calculate the Moore-Penrose pseudo-inversion \citep{pinv_penrose_1955} of the covariance.
In performing the pseudo-inversion, we choose a conservative cutoff of $10^{-15}$ on eigenvalues, but the results are not sensitive to the specific choice. 
We do not account for the Hartlap factor \citep{hartlap2007} in the inversion, as the effective number of data points is not well defined due to the correlation between the data bins.
This choice does not affect our parameter constraints, but may have a moderate effect on the absolute values of $\chi^2$.
In addition, the covariance matrix itself depends upon the parameters of the model; however, we neglect this dependence as covariances tend to be relatively weakly varying functions of the model parameters \citep{eifler_etal2009}. 
We assume uniform priors within certain intervals on the parameters, listed in \autoref{tab:priors}.

\begin{table}
    \centering
    \caption{\textbf{Prior intervals adopted for MCMC fits.}
    The $\Acen$ and $\Asat$ priors only apply to the decorated HOD fits, whereas in the standard HOD model both parameters are fixed to 0.}
    \label{tab:priors}
    \begin{tabular}{c|c}
    \hline\hline
    Parameter & Prior interval \\
    \hline
    $\log \Mmin$ & [11.0, 14.0] \\
    $\slogM$ & [0.02, 1.5] \\
    $\alpha$ &  [0, 1.5] \\
    $\log \Mone$ & [11.5, 15.0] \\
    $\log \Mzero$ & [9.0, 14.0] \\
    $\Acen$ & [-1, 1] \\
    $\Asat$ & [-1, 1] \\
    \hline\hline
    \end{tabular}
\end{table}

With the above likelihood and prior, we use the \texttt{emcee} \citep{emcee} package, which implements an affine-invariant ensemble sampler \citep{goodman_weare10}, to sample from the posterior distribution in the parameter space.
We use the \texttt{TabCorr}\footnote{Available at \url{https://github.com/johannesulf/TabCorr}.} package, which is based on the tabulating method proposed in \citet{zheng_guo2016_tabcorr}, to expedite the computation of $\wprp$.

\section{Results}
\label{sec:results}

In this section, we report the constraints we get on the galaxy–halo connection, making comparisons between the different statistic sets, models, and samples.

The galaxy samples we analyse are selected from SDSS DR7.
These include the five main samples listed in \autoref{tab:sample}, that maximise sample depths for the $M_r<-21.0$, $M_r<-20.5$, $M_r<-20.0$, $M_r<-19.5$, and $M_r<-19.0$ luminosity thresholds.
The results for the alternative $M_r<-20.0$ and $M_r<-19.5$ samples that exclude the Sloan Great Wall will be shown in \autoref{sec:alt_samples}.
For each SDSS galaxy sample, we consider four fitting cases:
\begin{enumerate}
    \item the standard HOD inferred from $\ngal+\wprp$;
    \item the standard HOD inferred from $\ngal+\wprp+\Pncic$;
    \item the decorated HOD inferred from $\ngal+\wprp$; and
    \item the decorated HOD inferred from $\ngal+\wprp+\Pncic$.
\end{enumerate}

\subsection{Marginalised Constraints on Assembly Bias Parameters}
\label{sec:AB_marginalised}

We first present the 1-D marginalised constraints on each HOD and dHOD parameter for the five main samples and the four fitting cases.
In \autoref{tab:fit_values}, we list the inferred parameter values, quoting the medians of the marginalised distributions as central values, with asymmetric ``1$\sigma$'' errors that correspond to the 16th and 84th percentile bounds. 

Our primary scientific focus is on the assembly bias parameters and we therefore begin our discussion with those parameters.
As is evident from the values in the seventh column of \autoref{tab:fit_values}, there are several samples for which clustering data suggest positive central galaxy assembly bias and for which the addition of CIC data provide further support for positive central galaxy assembly bias.
For the $M_r<-20.5$, $M_r<-19.5$, and $M_r<-19.0$ samples, the probability that $\Acen\le0$ is $3.0\%$, $0.1\%$, and $6.4\%$ respectively.
For the $M_r<-20.0$ sample, cases with $\Acen\le0$ are not detected with the current length of the chain, and the probability that $\Acen\le0.3$ is $0.1\%$.
Positive values of $\Acen$ correspond to the case in which galaxies are more likely to reside within high-concentration haloes and our analysis reveals intriguing indications that this may be the case for several of the SDSS DR7 luminosity threshold samples. 
In contrast to central galaxy assembly bias, there is no strong evidence of satellite galaxy assembly bias. 

\begin{figure}
    \centering
    \includegraphics[scale=0.43]{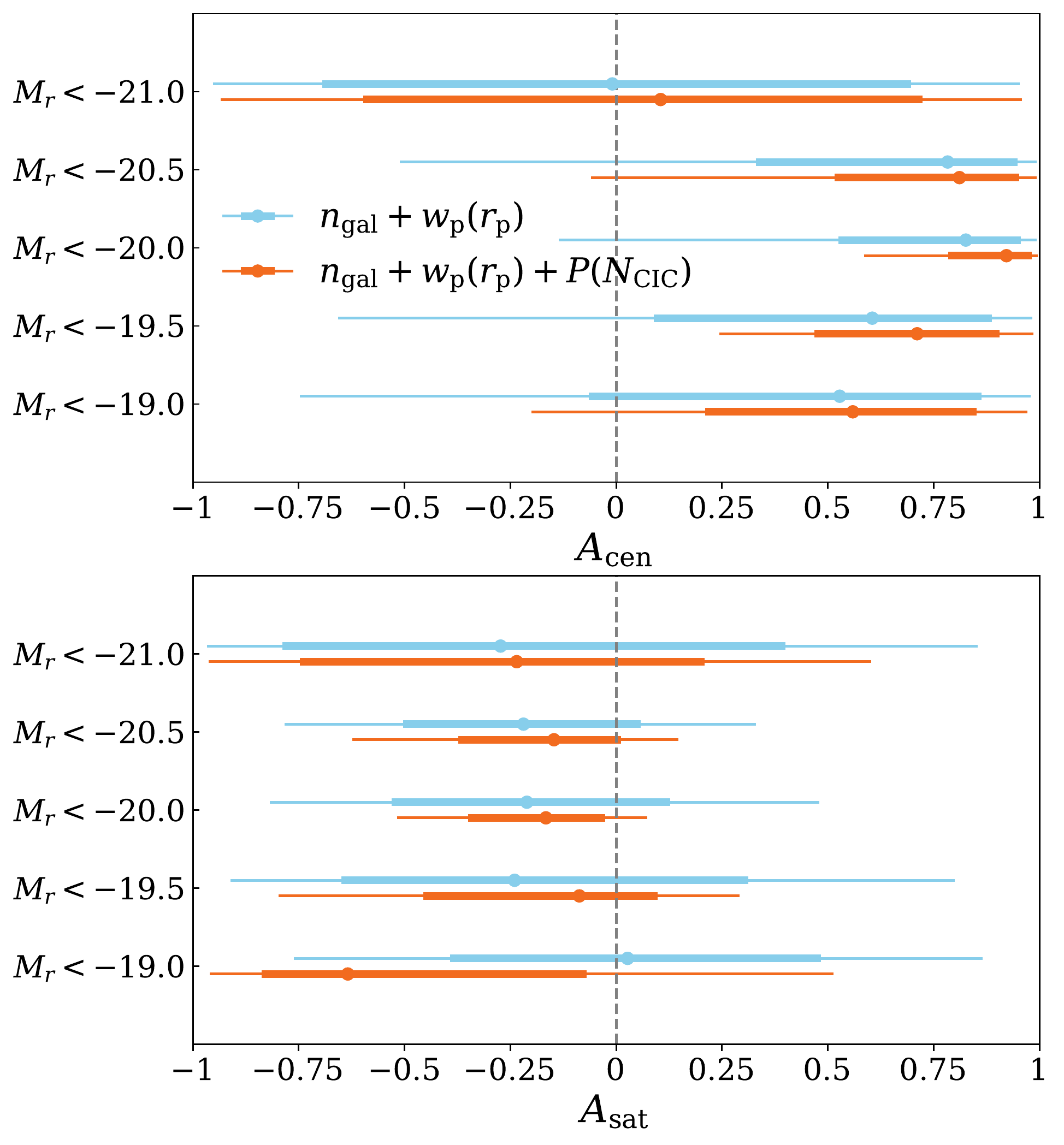}
    \caption{\textbf{Posterior 1$\sigma$ and 2$\sigma$ intervals of galaxy assembly bias parameters.}
    The two panels show the 1$\sigma$ (thick error bars) and 2$\sigma$ (thin error bars) constraints on the central and satellite assembly bias parameters, $\Acen$ and $\Asat$, respectively.
    The statistic sets from which the constraints are derived are shown in different colours ($\ngal+\wprp$ in blue; $\ngal+\wprp+\Pncic$ in orange). The different samples are laid out vertically and labelled on the $y$-axis.
    The markers show the medians in the marginalised posterior distributions, the thick error bars span between the 16th and 84th percentiles, and the thin error bars span between the 2.5th and 97.5th percentiles.
    The zero point, at which there is no assembly bias, is marked with a gray dashed line.}
    \label{fig:Agal_12s}
\end{figure}

We highlight the constraints on the assembly bias parameters in \autoref{fig:Agal_12s}. 
1$\sigma$ and 2$\sigma$ error bars are shown for the central and satellite assembly bias parameters, $\Acen$ and $\Asat$, separately, in the two panels.
For each parameter and each galaxy sample, the figure compares the constraints from the  clustering-only fits (``$\ngal$+$\wprp$'') and the fits that also include CIC (``$\ngal$+$\wprp$+$\Pncic$'').
Notice that the addition of companion counts in cylinders not only significantly reduces the confidence ranges on the inferred assembly bias parameters, $\Acen$ and $\Asat$, but also bolsters the evidence for $\Acen>0$ for all threshold samples. 
This in turn strengthens our ability to distinguish between the standard mass-only HOD and the decorated HOD that includes galaxy assembly bias.
We will discuss this result in more detail, in \autoref{sec:gab_detection} and \autoref{sec:model_compare}.

\begin{table*}[]
\caption{\textbf{Inferred constraints on HOD and dHOD parameters.}
For each sample, we tabulate the constraints on the parameters for the four cases listed in \autoref{sec:results}.
We quote the median of the marginalised distribution of each parameter as our fitting result, and the bounds of the asymmetric error bars correspond to the 16\% and 84\% percentiles.
There are five parameters for the HOD model, and seven for the decorated HOD.
$\log\Mmin$ and $\slogM$ are parameters that control the central galaxy occupation, and $\alpha$, $\log\Mone$ and $\log\Mzero$ control the satellite occupation.
$\Acen$ and $\Asat$ are the galaxy assembly bias parameters for centrals and satellites respectively, and are only applicable to the decorated HOD model.
$\Mmin$ is the mass at which a halo has a 50\% probability to host a central galaxy; 
$\slogM$ determines the steepness of the central occupation's transition from zero 
to unity; 
$\alpha$ is the index of the satellite occupation power law;
$\Mone$ indicates the halo mass at which the mean satellite number is one;
and $\Mzero$ is the mass below which there can be no centrals.
$\Acen$ and $\Asat$ can vary independently of each other, both ranging from -1 to 1, and in the absence of assembly bias, they assume values of 0.
}
\label{tab:fit_values}
\begin{tabular}{llllllll}
\hline\hline
$M_r<-21.0$                                                       & $\log\Mmin$                & $\slogM$                  & $\alpha$                  & $\log\Mone$                & $\log\Mzero$               & $\Acen$                    & $\Asat$                    \\ \hline
\begin{tabular}[c]{@{}l@{}}$\ngal+\wp$\\ HOD\end{tabular}         & $12.730^{+0.113}_{-0.064}$ & $0.358^{+0.212}_{-0.232}$ & $1.142^{+0.102}_{-0.210}$ & $13.945^{+0.054}_{-0.134}$ & $12.622^{+0.508}_{-2.144}$ & -- --                         & -- --                         \\
\begin{tabular}[c]{@{}l@{}}$\ngal+\wp+\Pncic$\\ HOD\end{tabular}  & $12.679^{+0.059}_{-0.027}$ & $0.228^{+0.164}_{-0.157}$ & $0.938^{+0.086}_{-0.092}$ & $13.817^{+0.066}_{-0.075}$ & $13.113^{+0.131}_{-0.180}$ & -- --                         & -- --                         \\
\begin{tabular}[c]{@{}l@{}}$\ngal+\wp$\\ dHOD\end{tabular}        & $12.716^{+0.123}_{-0.054}$ & $0.318^{+0.242}_{-0.209}$ & $1.103^{+0.115}_{-0.188}$ & $13.979^{+0.049}_{-0.118}$ & $12.469^{+0.589}_{-2.130}$ & $-0.009^{+0.705}_{-0.686}$ & $-0.273^{+0.673}_{-0.515}$ \\
\begin{tabular}[c]{@{}l@{}}$\ngal+\wp+\Pncic$\\ dHOD\end{tabular} & $12.671^{+0.063}_{-0.020}$ & $0.185^{+0.198}_{-0.123}$ & $0.892^{+0.093}_{-0.094}$ & $13.844^{+0.091}_{-0.088}$ & $13.086^{+0.152}_{-0.202}$ & $0.105^{+0.619}_{-0.702}$  & $-0.235^{+0.444}_{-0.512}$ \\ \hline
$M_r<-20.5$                                                       & $\log\Mmin$                & $\slogM$                  & $\alpha$                  & $\log\Mone$                & $\log\Mzero$               & $\Acen$                   & $\Asat$                    \\ \hline
\begin{tabular}[c]{@{}l@{}}$\ngal+\wp$\\ HOD\end{tabular}         & $12.258^{+0.049}_{-0.023}$ & $0.172^{+0.172}_{-0.108}$ & $1.095^{+0.061}_{-0.069}$ & $13.544^{+0.050}_{-0.059}$ & $12.264^{+0.289}_{-1.223}$ & -- --                        & -- --                         \\
\begin{tabular}[c]{@{}l@{}}$\ngal+\wp+\Pncic$\\ HOD\end{tabular}  & $12.246^{+0.016}_{-0.011}$ & $0.105^{+0.085}_{-0.059}$ & $0.947^{+0.033}_{-0.034}$ & $13.432^{+0.033}_{-0.036}$ & $12.681^{+0.078}_{-0.089}$ & -- --                        & -- --                         \\
\begin{tabular}[c]{@{}l@{}}$\ngal+\wp$\\ dHOD\end{tabular}        & $12.325^{+0.166}_{-0.075}$ & $0.402^{+0.278}_{-0.254}$ & $1.041^{+0.070}_{-0.073}$ & $13.528^{+0.054}_{-0.065}$ & $12.267^{+0.305}_{-1.053}$ & $0.783^{+0.165}_{-0.453}$ & $-0.219^{+0.277}_{-0.284}$ \\
\begin{tabular}[c]{@{}l@{}}$\ngal+\wp+\Pncic$\\ dHOD\end{tabular} & $12.265^{+0.031}_{-0.021}$ & $0.218^{+0.099}_{-0.104}$ & $0.950^{+0.042}_{-0.045}$ & $13.454^{+0.039}_{-0.038}$ & $12.617^{+0.096}_{-0.125}$ & $0.811^{+0.141}_{-0.295}$ & $-0.147^{+0.158}_{-0.226}$ \\ \hline
$M_r<-20.0$                                                       & $\log\Mmin$                & $\slogM$                  & $\alpha$                  & $\log\Mone$                & $\log\Mzero$               & $\Acen$                   & $\Asat$                    \\ \hline
\begin{tabular}[c]{@{}l@{}}$\ngal+\wp$\\ HOD\end{tabular}         & $11.953^{+0.091}_{-0.032}$ & $0.238^{+0.243}_{-0.150}$ & $1.018^{+0.063}_{-0.065}$ & $13.180^{+0.073}_{-0.081}$ & $12.393^{+0.234}_{-0.398}$ & -- --                        & -- --                         \\
\begin{tabular}[c]{@{}l@{}}$\ngal+\wp+\Pncic$\\ HOD\end{tabular}  & $11.937^{+0.018}_{-0.014}$ & $0.109^{+0.094}_{-0.066}$ & $0.902^{+0.031}_{-0.032}$ & $13.072^{+0.040}_{-0.046}$ & $12.552^{+0.085}_{-0.088}$ & -- --                        & -- --                         \\
\begin{tabular}[c]{@{}l@{}}$\ngal+\wp$\\ dHOD\end{tabular}        & $12.167^{+0.401}_{-0.204}$ & $0.690^{+0.450}_{-0.396}$ & $0.955^{+0.069}_{-0.080}$ & $13.144^{+0.081}_{-0.110}$ & $12.382^{+0.288}_{-0.676}$ & $0.826^{+0.130}_{-0.301}$ & $-0.211^{+0.339}_{-0.319}$ \\
\begin{tabular}[c]{@{}l@{}}$\ngal+\wp+\Pncic$\\ dHOD\end{tabular} & $11.973^{+0.027}_{-0.020}$ & $0.298^{+0.058}_{-0.062}$ & $0.928^{+0.033}_{-0.039}$ & $13.118^{+0.043}_{-0.050}$ & $12.420^{+0.113}_{-0.123}$ & $0.922^{+0.060}_{-0.138}$ & $-0.166^{+0.140}_{-0.184}$ \\ \hline
$M_r<-19.5$                                                       & $\log\Mmin$                & $\slogM$                  & $\alpha$                  & $\log\Mone$                & $\log\Mzero$               & $\Acen$                   & $\Asat$                    \\ \hline
\begin{tabular}[c]{@{}l@{}}$\ngal+\wp$\\ HOD\end{tabular}         & $11.806^{+0.453}_{-0.144}$ & $0.547^{+0.592}_{-0.389}$ & $1.049^{+0.041}_{-0.058}$ & $12.975^{+0.047}_{-0.077}$ & $11.599^{+0.677}_{-1.703}$ & -- --                     & -- --                      \\
\begin{tabular}[c]{@{}l@{}}$\ngal+\wp+\Pncic$\\ HOD\end{tabular}  & $11.751^{+0.034}_{-0.032}$ & $0.443^{+0.069}_{-0.076}$ & $0.913^{+0.033}_{-0.036}$ & $12.817^{+0.048}_{-0.057}$ & $12.449^{+0.107}_{-0.120}$ & -- --                     & -- --                      \\
\begin{tabular}[c]{@{}l@{}}$\ngal+\wp$\\ dHOD\end{tabular}        & $11.940^{+0.432}_{-0.262}$ & $0.762^{+0.510}_{-0.530}$ & $0.993^{+0.065}_{-0.071}$ & $12.947^{+0.078}_{-0.098}$ & $11.655^{+0.676}_{-1.669}$ & $0.605^{+0.282}_{-0.516}$ & $-0.240^{+0.552}_{-0.409}$ \\
\begin{tabular}[c]{@{}l@{}}$\ngal+\wp+\Pncic$\\ dHOD\end{tabular} & $11.709^{+0.029}_{-0.023}$ & $0.367^{+0.066}_{-0.064}$ & $0.890^{+0.040}_{-0.049}$ & $12.785^{+0.054}_{-0.059}$ & $12.527^{+0.100}_{-0.111}$ & $0.711^{+0.194}_{-0.243}$ & $-0.087^{+0.185}_{-0.369}$ \\ \hline
$M_r<-19.0$                                                       & $\log\Mmin$                & $\slogM$                  & $\alpha$                  & $\log\Mone$                & $\log\Mzero$               & $\Acen$                   & $\Asat$                    \\ \hline
\begin{tabular}[c]{@{}l@{}}$\ngal+\wp$\\ HOD\end{tabular}         & $11.733^{+0.395}_{-0.158}$ & $0.594^{+0.550}_{-0.404}$ & $1.046^{+0.031}_{-0.036}$ & $12.838^{+0.036}_{-0.044}$ & $10.477^{+1.064}_{-1.005}$ & -- --                        & -- --                         \\
\begin{tabular}[c]{@{}l@{}}$\ngal+\wp+\Pncic$\\ HOD\end{tabular}  & $11.597^{+0.041}_{-0.048}$ & $0.410^{+0.087}_{-0.149}$ & $0.894^{+0.050}_{-0.042}$ & $12.649^{+0.071}_{-0.072}$ & $12.328^{+0.130}_{-0.185}$ & -- --                        & -- --                         \\
\begin{tabular}[c]{@{}l@{}}$\ngal+\wp$\\ dHOD\end{tabular}        & $11.797^{+0.390}_{-0.218}$ & $0.710^{+0.525}_{-0.477}$ & $1.028^{+0.038}_{-0.049}$ & $12.840^{+0.061}_{-0.062}$ & $10.541^{+1.073}_{-1.048}$ & $0.528^{+0.335}_{-0.593}$ & $0.027^{+0.457}_{-0.420}$  \\
\begin{tabular}[c]{@{}l@{}}$\ngal+\wp+\Pncic$\\ dHOD\end{tabular} & $11.565^{+0.039}_{-0.031}$ & $0.356^{+0.102}_{-0.114}$ & $0.842^{+0.047}_{-0.045}$ & $12.649^{+0.065}_{-0.070}$ & $12.316^{+0.121}_{-0.135}$ & $0.559^{+0.292}_{-0.349}$ & $-0.634^{+0.564}_{-0.203}$ \\ \hline\hline
\end{tabular}
\end{table*}

\subsection{Full Posterior of Model Parameters}
\label{sec:full_post}

\begin{figure*}
    \centering
    \includegraphics[scale=0.9, trim={0 1cm 0 1cm},clip]{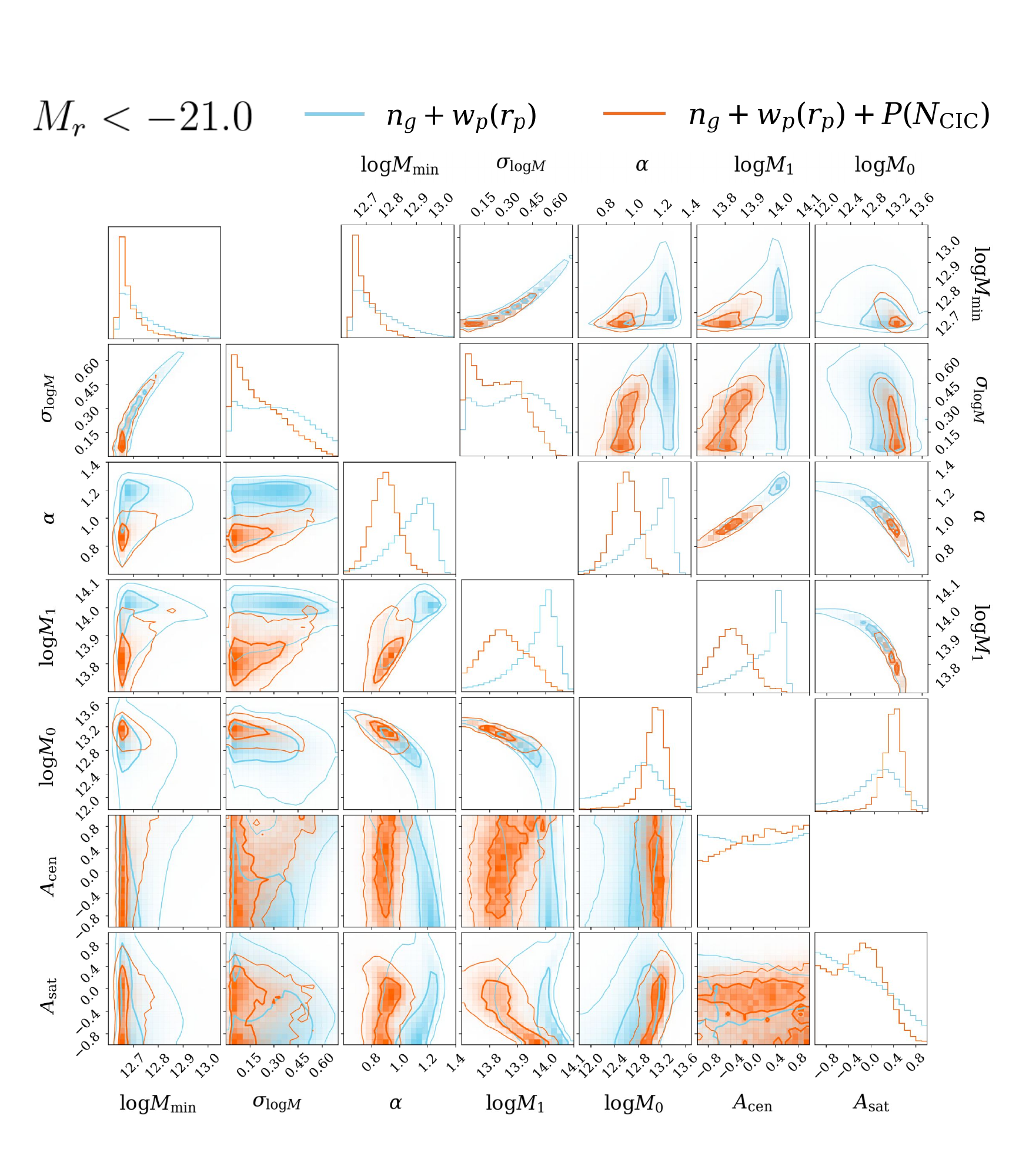}
    \caption{\textbf{The standard HOD and decorated HOD fits for the $M_r<-21.0$ sample.}
    Fits for the standard HOD are shown in the upper right triangle, and fits for the decorated HOD, which has two additional assembly bias parameter, are shown in the lower left triangle.
    The contours show the $1\sigma$ and $2\sigma$ constraints as two-dimensional projections in the parameter space, where the $1\sigma$ contours are plotted in thicker lines.
    The one-dimensional histograms are marginalised for each individual parameter.
    The two sets of contours with different colours show the constraints from $\ngal$ and $\wprp$ (blue), and the constraints from $\ngal$, $\wprp$ and $\Pncic$ combined (orange), respectively.
    The discontinuities in the contours are plotting artefacts and do not reflect irregularities in the constraints.}
    \label{fig:corner21p0}
\end{figure*}

\begin{figure*}
    \centering
    \includegraphics[scale=0.9, trim={0 1cm 0 1cm},clip]{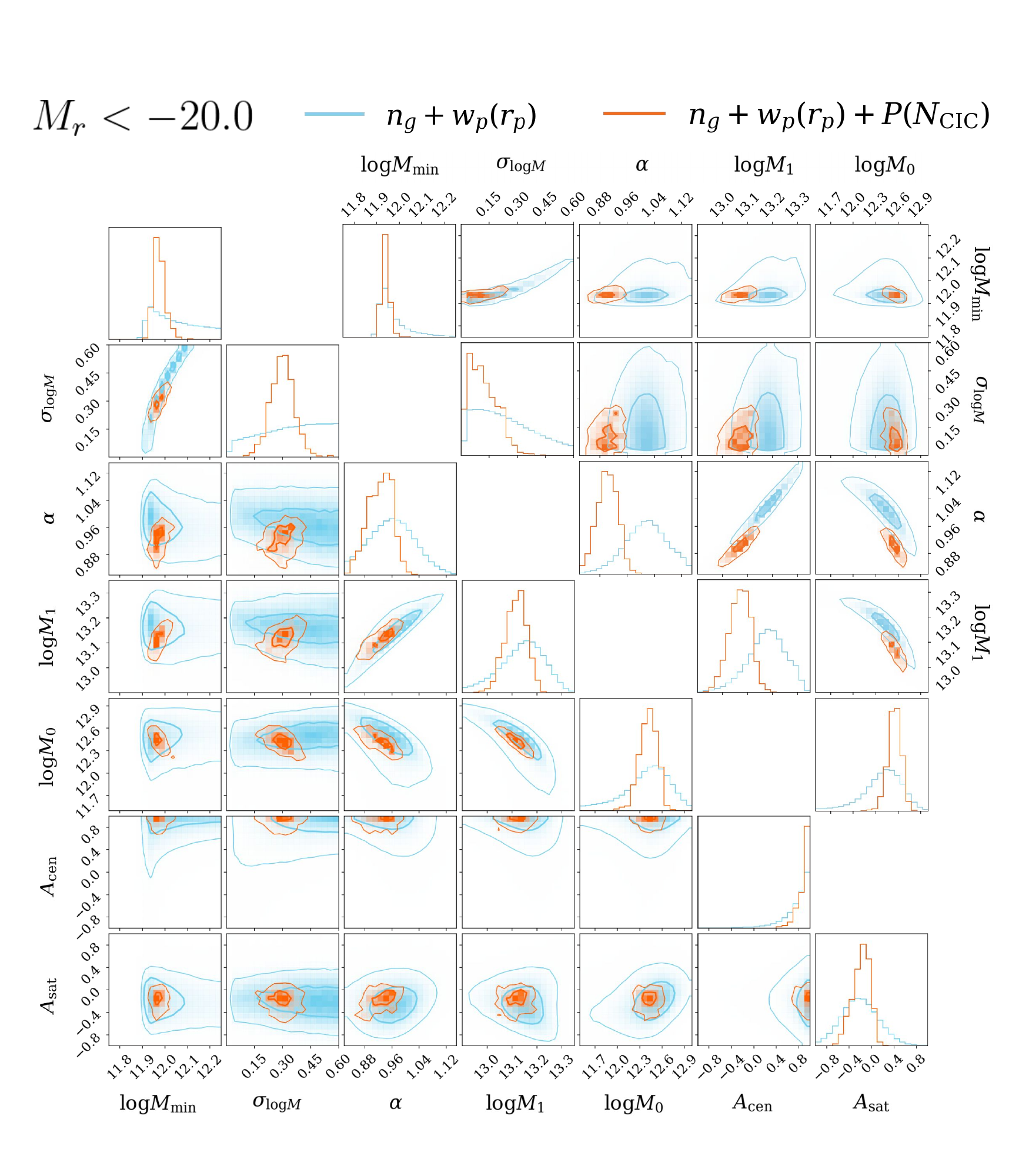}
    \caption{Same as \autoref{fig:corner21p0}, but for the $M_r<-20.0$ sample.}
    \label{fig:corner20p0}
\end{figure*}

\begin{figure*}
    \centering
    \includegraphics[scale=0.9, trim={0 1cm 0 1cm},clip]{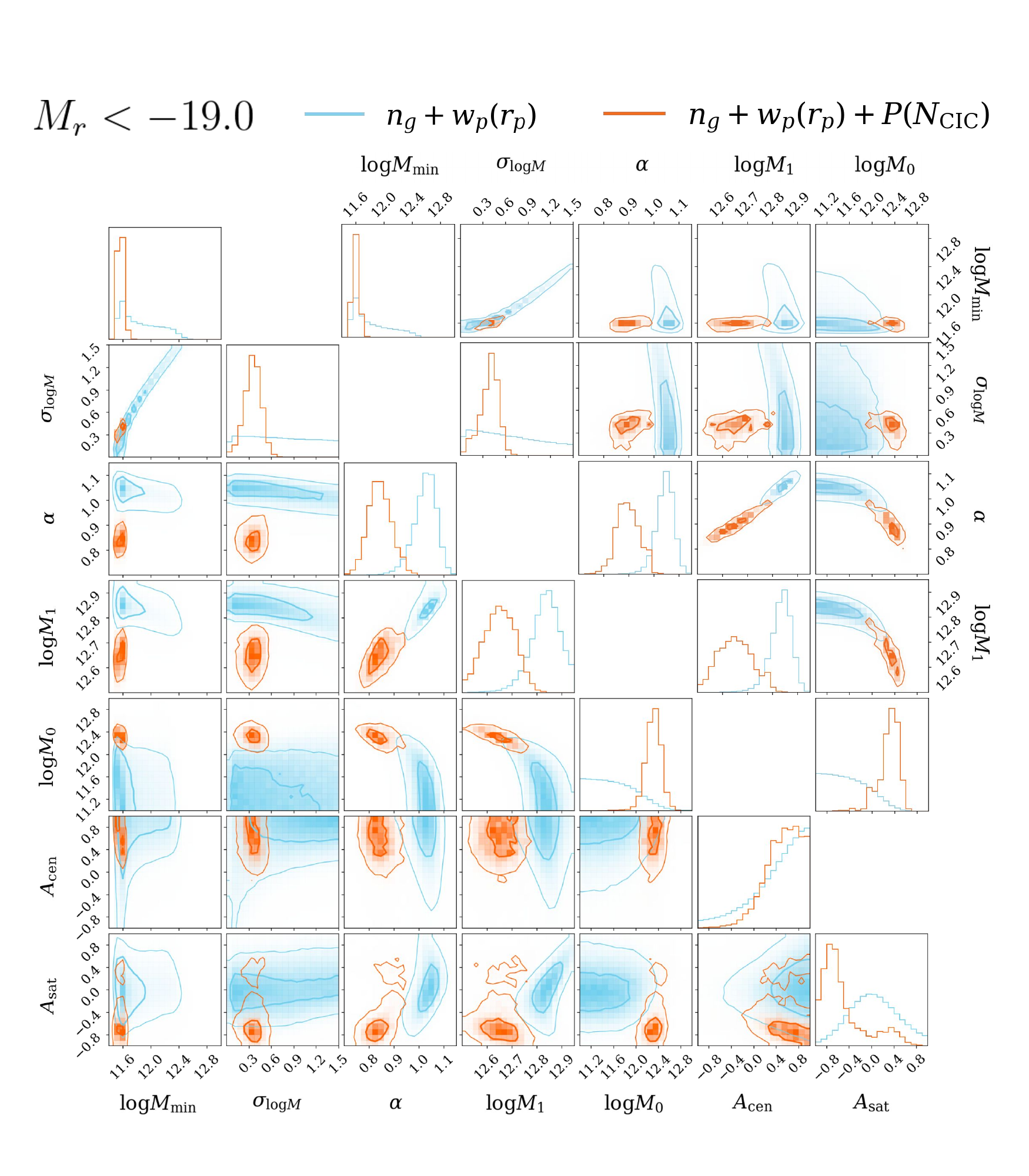}
    \caption{Same as \autoref{fig:corner21p0}, but for the $M_r<-19.0$ sample.}
    \label{fig:corner19p0}
\end{figure*}

In addition to the constraints listed in 
\autoref{tab:fit_values}, we also depict the full posterior of all model parameters visually in a series of figures. 
\autoref{fig:corner21p0} shows results for the $M_r<-21$ sample, \autoref{fig:corner20p0} shows results for the $M_r<-20$ sample, and \autoref{fig:corner19p0} gives the $M_r<-19$ results.
We do not show two-dimensional constraints for $M_r<-20.5$ and $M_r<-19.5$ in the interest of brevity, since they do not add any qualitatively new information.
The figures show the 1$\sigma$ and 2$\sigma$ marginalised constraints in two dimensions for all pairwise combinations of parameters.
In each of these figures, constraints on the five parameters of the HOD are shown in the upper right triangle, and constraints on the seven parameters of the dHOD are exhibited in the lower right triangle.
In each triangle, each diagonal panel shows the entire one-dimensional marginalised posterior distribution for the parameter listed for the column.
Furthermore, each panel shows results from fits to clustering only (``$\ngal$+$\wprp$'') as well as results from the simultaneous fits to clustering and CIC distributions (``$\ngal$+$\wprp$+$\Pncic$''), in different colours.

There are several points to note from these figures. In each of the fits to clustering (``$\ngal$+$\wprp$''), there are prominent degeneracies among the parameters.
For example, $\slogM$ and $\log\Mmin$ are degenerate, as are $\alpha$ and $\log\Mone$.
These degeneracies are well known and have been discussed in the earlier literature \citep[e.g.][]{zentner_etal19}.
As suggested by our earlier analysis in W19, the addition of $\Pncic$ to the set of observables partly breaks these degeneracies, resulting in considerably more restrictive parameter constraints.
In many cases, the improvement is dramatic.
For example, the width of the marginalised 1$\sigma$ constraint on $\log\Mmin$ is reduced by a factor of $\sim 4$ for the HOD analysis of the $M_r<-20$ sample and by a factor of nearly 13 for the dHOD analysis of the same sample. 

Interestingly, the $M_r<-19.0$ sample shows a bimodality in the inferred posterior distribution of $\Asat$ from the $\ngal+\wprp+\Pncic$ analysis.
The effect of non-zero satellite assembly bias is to distribute more satellite galaxies into a subpopulation of haloes, which enhances the small-scale clustering regardless of the sign of $\Asat$.
In the jargon of the HOD model, this is an enhancement of the ``one-halo'' clustering. 
This small-scale, one-halo clustering is enhanced regardless of the sign of $\Asat$ (regardless of whether or not low-concentration or high-concentration haloes preferentially host satellite galaxies) because placing a fixed number of galaxies into fewer haloes increases the number of pairs among these galaxies. 
Consequently, both $\Asat > 0$ and $\Asat < 0$ yield enhanced small-scale clustering, and it is not at all surprising to find that when values of $\Asat \ne 0$ are preferred by data, there is significant probability in the posterior distribution corresponding to both $\Asat < 0$ and $\Asat > 0$. 
The enhancement is not perfectly symmetric between positive and negative values of $\Asat$ for a variety of reasons \citep{hearin_etal16}.
Moreover, this degeneracy is partially broken by the addition of $\Pncic$, which contains higher-order information of the small-scale galaxy distribution, and is able to better distinguish between positive, zero, and negative values of $\Asat$. 

Notice additionally, that there are some interesting notes of tension among our analyses. 
In the HOD analysis of the $M_r<-20$ sample (\autoref{fig:corner20p0}, upper right), there is some weak tension between the clustering-only fits (``$\ngal$+$\wprp$'') and the fits that include both clustering and CIC (``$\ngal$+$\wprp$+$\Pncic$'').
While this tension is weak, it is also mitigated in the dHOD fit to the same sample, which includes assembly bias (\autoref{fig:corner20p0}, lower left).
More interesting in this regard may be the $M_r<-19$ sample.
The HOD analysis of the $M_r<-19$ sample exhibits more sizeable tensions (\autoref{fig:corner19p0}, upper right).
Furthermore, this tension persists in the dHOD analysis which includes assembly bias (\autoref{fig:corner19p0}, lower left).
We discuss this tension further in \autoref{sec:discussion-tension}.

\section{Interpretation}
\label{sec:interpretation}

The initial motivation of this work was to leverage the degeneracy breaking power of CIC statistics to complement the two-point function in order to constrain and/or detect galaxy assembly bias.
The results of the previous section demonstrate that we have achieved this goal.
Our current task is now to interpret these results.
In this section, we examine the degree to which our results may be interpreted as a detection of assembly bias, the comparison between the standard HOD and the decorated HOD results, and the trends among different luminosity threshold samples.

\subsection{Detection of Galaxy Assembly Bias}
\label{sec:gab_detection}

We begin with a discussion of the degree to which our results may be interpreted as a detection of assembly bias in the context of the dHOD model.
In general, the data prefer positive $\Acen$, indicating that central galaxies of each luminosity threshold are more likely to reside within high-concentration haloes.
In several samples, $\Acen \le 0$ is ruled out at relatively high significance.
We cast the strength of these assembly bias detections into approximate ``sigma level measures,'' using the method described in Section 2.2 of \citet{Kohlinger2019}. 
Namely, we derive posterior distributions of the differences between the parameters in the chains and parameter instances without assembly bias, quantify the rareness of zero difference, and convert it to the sigma level according to the probability distribution of a one-dimensional Gaussian.
Strong levels of positive central assembly bias are observed in the $M_r<-20.0$ (at least 4.4$\sigma)$ and $M_r<-19.5\ (3.4\sigma)$ samples.
Positive central assembly bias is also preferred by the $M_r<-20.5\ (2.2\sigma)$ and $M_r<-19.0\ (1.8\sigma)$ samples.
We find no evidence of central galaxy assembly bias in the brightest $M_r<-21.0$ sample, from which constraints are generally quite weak due to the larger statistical errors on this sample.
In comparison, we find only relatively weak indications that $\Asat$ may be negative. 
Non-zero satellite assembly bias is only marginally suggested in the $M_r<-20.0\ (1.0\sigma)$ and $M_r<-19.0\ (1.4\sigma)$ samples, and not suggested by data from the other samples.

The senses of these signals are consistent with our expectations.
At a fixed halo mass, haloes with deeper potential wells and thus higher concentrations are more likely to host central galaxies.
On the other hand, haloes with higher concentrations tend to have less surviving substructure, and therefore fewer satellite galaxies.
We discuss the physical origin of our signals in more detail in \autoref{sec:hydro-sims}, referring to predictions of hydrodynamical simulations.

\subsection{Information Criteria of HOD and dHOD Models}
\label{sec:model_compare}

The preceding discussion presents the case that this analysis does represent a detection of assembly bias.
However, it is necessary to explore the degree to which the data prefer the dHOD model to the standard HOD model, as the decorated HOD (dHOD) is an extension to the standard HOD model. 

We quantify our model selection using the Akaike Information Criterion \citep[AIC,][]{AIC1973} and the Bayesian Information Criterion \cite[BIC,][]{schwarz78}, both of which penalise models for having more parameters. The information criteria are 
\begin{eqnarray}
{\rm AIC} = 2k - 2\ln \mathcal{L}_{\rm max},\\
{\rm BIC} = k\ln n - 2\ln \mathcal{L}_{\rm max},
\end{eqnarray}
where $k$ is the number of parameters in the model, $n$ is the number of data points involved in the fit, and $\mathcal{L}_{\rm max}$ the maximum likelihood, in our case $-2\ln \mathcal{L}_{\rm max}=\chi^2_{\rm min}$.
The model with the smaller AIC or BIC value is considered a better description of the data.
Generally, an extended model is considered to be strongly preferred if it reduces the AIC or BIC by 5 or more \citep[e.g.,][]{burnham2011aic}.

The information criteria, along with the minimum $\chi^2$ values normalised by the number of degrees of freedom, are listed in \autoref{tab:fit_stats}.
Our choice not to incorporate the Hartlap factor (see \autoref{sec:fitting}) may have a moderate effect on the absolute values of $\chi^2$ and the information criteria, but does not affect the comparison between models.
Comparing between the rows of HOD and dHOD fits for each sample, 
we find mixed results.
The fits in which clustering alone is used to constrain the galaxy-halo connection models (``$\ngal+\wprp$'') do not prefer the dHOD over HOD and even give a marginal preference to the standard HOD relative to the dHOD. 

Including CIC statistics in the analysis (``$\ngal+\wprp+\Pncic$'' analyses) also yields mixed conclusions. 
In these analyses, the dHOD model is often preferred. 
In fact, this preference is categorised as strong in the $M_r<-20$ and $M_r<-19.5$ samples. 
However, the opposite is true in the brightest sample. 
Our analysis of the $M_r<-21$ sample shows a preference for the HOD over the dHOD. 
It may be the case that assembly bias is not indicated in the most luminous sample because concentration-dependent halo clustering weakens as halo mass increases.
Therefore, a concentration-dependent halo occupation model induces only small changes in predicted galaxy clustering statistics in high-luminosity (i.e., high-mass) samples.

\begin{table*}[]
\caption{\textbf{Model selection and information criteria.}
For each galaxy sample, we list several measures for assessing the effectiveness of the HOD and dHOD models to fit the data measurements.
The measures include the $\chi^2$ value normalised by the degree of freedom, the Akaike Information Criterion, and the Bayesian Information Criterion.
These are calculated for both sets of statistics, $\ngal+\wprp$, and $\ngal+\wprp+\Pncic$.
We also list the level of tension between the parameters inferred from the two statistic sets, as a proxy for how well the model can fit all the statistics at once.}
\begin{tabular}{l|lllllllll}
\hline\hline
                  & \multicolumn{3}{l}{\hspace{23pt}$\ngal+\wprp$} & & \multicolumn{3}{l}{\hspace{1pt}$\ngal+\wprp+\Pncic$} && \multirow{2}{*}{Tension} \\
                  & $\chi^2$/DoF   & AIC     & BIC    && $\chi^2$/DoF     & AIC       & BIC       &&                          \\ \hline
$M_r<-21.0$, HOD  & 0.71                & 15.65      & 18.47      && 0.38                & 19.15      & 25.99      && 0.20$\sigma$                 \\
\hspace{40pt} dHOD & 0.69                & 18.11      & 22.07      && 0.42                & 23.19      & 32.76      && 0.07$\sigma$                 \\ \hline
$M_r<-20.5$, HOD  & 1.38                & 21.06      & 23.88      && 0.81                & 34.40      & 42.18      && 0.93$\sigma$                 \\
\hspace{40pt} dHOD & 1.25                & 21.49      & 25.45      && 0.62                & 31.35      & 42.24      && 0.18$\sigma$                 \\ \hline
$M_r<-20.0$, HOD  & 1.49                & 21.91      & 24.74      && 1.54                & 62.49      & 70.81      && 0.89$\sigma$                 \\
\hspace{40pt} dHOD & 1.18                & 21.08      & 25.04      && 1.05                & 47.69      & 59.34      && 0.07$\sigma$                 \\ \hline
$M_r<-19.5$, HOD  & 0.60                & 14.78      & 17.60      && 1.07                & 48.35      & 56.92      && 0.87$\sigma$                 \\
\hspace{40pt} dHOD & 0.57                & 17.40      & 21.36      && 0.81                & 41.63      & 53.61      && 0.25$\sigma$                 \\ \hline
$M_r<-19.0$, HOD  & 0.77                & 16.17      & 19.00      && 2.09                & 87.39      & 96.08      && 2.00$\sigma$                 \\
\hspace{40pt} dHOD & 0.92                & 19.51      & 23.47      && 1.99                & 83.65      & 95.82      && 1.84$\sigma$                 \\ \hline\hline
\end{tabular}
\label{tab:fit_stats}
\end{table*}

\subsection{Tension Between Analyses}
\label{sec:discussion-tension}

When constraining a model using different data sets, the results could be in tension with each other.
In other words, the model is unable to simultaneously reproduce the different data, which could indicate incompleteness in the modelling choices.
The consistency between constraints from $\ngal+\wprp$ and $\ngal+\wprp+\Pncic$ thus provides a test of the model's ability of describing the real Universe.
We quantify this consistency in \autoref{sec:model_compare}, where we compare the performance of the HOD and the dHOD models.
A consequence of tensions between statistics is a decrease in the constraining power of the joint analysis, as the model struggles to fit both observables.
This could explain the behaviour of the $M_r<-19.0$ sample, a rare case for which the constraints on $\alpha$ and $\log\Mone$ worsen when $\Pncic$ is included, as the preferences for those parameters indeed show a significant discrepancy between the data sets.

Also in \autoref{tab:fit_stats}, we quantify the consistency between the two sets of parameter constraints, with and without $\Pncic$, again using the method in \citet{Kohlinger2019}.
As discussed previously, this measure indirectly reflects the ability of the models to describe the underlying physics.
In most of the samples, we find no significant tension between the data sets for either model, however, we note that the dHOD model shows a universally superior consistency between data sets.
For the $M_r<-19.0$ sample, there is a 2.0$\sigma$ tension between the observable sets, which is slightly alleviated (1.8$\sigma$) when including assembly bias.
The tension could be due to specific modelling choices, such as the simple form of the spatial profile and velocities of satellites in haloes, especially as $M_r<-19.0$ is the faintest sample and has the highest fraction of satellite galaxies.
It could also be due to the form of incorporation of galaxy assembly bias, or the choice to only categorise haloes into two discrete subpopulations by concentration.
Overall, the decorated HOD is a superior model for accounting for the different statistics at once, though there is still room for improvement in the modelling choices.

\subsection{Comparison Across Galaxy Samples}
\label{sec:sample_compare}

The different galaxy samples represent populations that have different luminosities, reside in different environments, and are observed in different quantities.
The model parameters, as well as the constraining power, show dependencies on the luminosity threshold.
Our conclusions regarding galaxy assembly bias are also different among the different samples.
We discuss the trends with luminosity thresholds in this subsection.

We first examine how parameter values vary with the luminosity threshold used for sample selection.
The three mass scale parameters, $\log\Mmin$, $\log\Mone$ and $\log\Mzero$, show a clear trend of increase when going to brighter samples.
This is consistent with our expectation that brighter galaxies live in more massive haloes.
Overall $\slogM$ becomes smaller for brighter samples, which suggests that the scatter in the stellar mass–halo mass relation tends to be smaller on the more massive end \citep[in agreement with][]{Lange.etal.19b}, but the trend is not very pronounced, and there are exceptions.
There isn't a clear and consistent trend in $\alpha$ with luminosity.

Another comparison can be made between the tightness of constraints for the different luminosity samples, which is not a monotonic function of luminosity.
Instead, the constraining power is lower on both the faint and the bright ends, and reaches a maximum for the intermediate samples.
This is generally true for the different parameters, and consistent with our expectations.
The brighter samples have lower number densities, and the errors of the statistics are dominated by shot noise, whereas the fainter samples are limited by flux, in which case cosmic variance due to finite volume becomes the main source of uncertainty.
The intermediate samples are not as severely susceptible to either effect as the more extreme samples, and the balance leads to the best constraints that we can obtain.
We also note the difference between the impact of small-scale shot noise on $\wprp$ and $\Pncic$ in the bright samples.
The constraining power of $\wprp$ comes both from smaller and larger radii, while $\Pncic$ focuses on the higher-order clustering at small scales, which causes $\Pncic$ to be even more sensitive to the decrease of galaxy numbers, and undermines its ability to put constraints on parameters.
This is indeed what we observe on the bright end of the samples: constraints from $\ngal+\wprp+\Pncic$ worsen much more than $\ngal+\wprp$ only.

An additional factor contributes to the luminosity dependence of the constraints on the assembly bias component of the model.
The manifestation of galaxy assembly bias in galaxy clustering relies on the existence of halo assembly bias, which is the dependence of halo clustering on secondary halo properties.
Strong levels of halo assembly bias therefore amplify the effect of galaxy assembly bias on galaxy clustering, allowing easier detection.
In the halo mass range relevant to our galaxy samples, the dependence of halo clustering strength on halo concentration weakens toward the high mass end \citep{wechsler06,gao_white07}, and further suppresses any galaxy assembly bias signal that may be present in the $M_r<-21.0$ sample.

\section{Discussion}
\label{sec:discussion}

In this section, we look into the physics behind our results, discuss our modelling choices, and compare our findings with previous work.

\subsection{Galaxy Assembly Bias in Hydrodynamical Simulations}
\label{sec:hydro-sims}

The results presented in the previous sections strongly suggest the existence of central galaxy assembly bias in the sense that central galaxies of a fixed luminosity are more likely to inhabit high-concentration haloes at fixed halo mass.
The results also provide a hint of satellite galaxy assembly bias, where galaxies of a fixed luminosity are more likely to inhabit low-concentration halos.
The senses of these assembly bias signals are in line with our gross physical intuition.
Central galaxies are more likely to form in haloes with deeper potential wells at fixed mass. 
These haloes correspond precisely to high-concentration haloes. 
Moreover, at a given halo mass, haloes that assembled earlier have higher concentrations \citep[e.g.,][]{nfw97} and less surviving substructure that may host satellite galaxies \citep[e.g.,][]{zentner_etal05,Mao2015,Jiang_vdB17}.
Correspondingly, low-concentration haloes tend to have more satellite haloes at fixed mass. 
Hence, both a positive $\Acen$ and a negative $\Asat$ are in line with our physical expectations.

There have been a number of studies of assembly bias within galaxy formation simulations.
The most recent of these is \citet{yuan_etal22}, who studied assembly bias in galaxy samples selected from the IllustrisTNG-300-1 simulation to mock the DESI LRG and ELG galaxy samples.
While neither of these samples is directly comparable to the SDSS DR7 Main Galaxy Sample, it is interesting to note that the LRG sample of galaxies exhibits galaxy assembly bias in the same sense as what we observe in our study.
In particular, central galaxy LRGs in IllustrisTNG preferentially inhabit haloes with concentrations that are higher than average while satellite LRGs preferentially inhabit haloes with lower concentrations.
On the other hand, both central and satellite ELGs are found preferentially within haloes with concentrations that are lower than average, a result that is not surprising because ELGs are thought to be preferentially newly forming galaxies that are found preferentially within newly forming haloes.

A number of other studies are worthy of note.
\citet{montero-dorta_etal21} studied assembly bias more generally within IllustrisTNG-300-1.
Broadly speaking, these authors found that galaxies with stellar masses above $10^7 \, \mathrm{M}_{\odot}$ preferentially inhabit early-forming haloes and as early-forming haloes are preferentially high-concentration haloes, this result is qualitatively consistent with the enhanced probability of central galaxies to reside in high-concentration host haloes at fixed mass which we infer from the SDSS data.
\citet{hadzhiyska_etal21} compared assembly bias present in the IllustrisTNG-300-1 simulation with that in the Santa Cruz semi-analytic galaxy formation model \citep{gabrielpillai_etal22}.
In both the simulation and the semi-analytic model, central galaxies preferentially reside within haloes of higher concentration at fixed halo mass while satellite galaxies are found with higher probability in lower-concentration haloes.
This is consistent with the earlier findings of \citet{Bose_2019} who studied galaxy clustering in IllustrisTNG-100-1 and IllustrisTNG-300-1 simulations.
\citet{xu_zheng2020} came to similar conclusions in their study of the Illustris galaxies and determined that most of the effect of assembly bias could be accounted for by tabulating the galaxy content of haloes as a function of peak circular velocity, rather than halo mass.
\citet{contreras2019} studied assembly bias in the semi-analytic galaxy formation model of \citet{guo_etal13}, finding results in qualitative agreement with the aforementioned studies.
Each of these recent studies are consistent with earlier work on the EAGLE simulation \citep{chaves-montero16}.
Interestingly, all of these results from studies of simulations and semi-analytic models are in qualitative agreement with the sense of assembly bias that we infer from our SDSS analyses. 

\subsection{Modelling of Galaxy Assembly Bias}

In this work, we obtain constraints on galaxy assembly bias in terms of the decorated HOD model, with halo concentration as the secondary halo property.
We make this choice for several reasons.

Of the commonly considered halo properties, halo concentration has been found to exhibit strong correlations with both halo assembly history \citep[e.g.,][]{wechsler02} and halo clustering \citep[e.g.,][]{wechsler06}.
The former is closely connected with galaxy formation, while the latter is a potential source of observable signals, rendering concentration likely to yield a detection of galaxy assembly bias.
Meanwhile, other secondary halo properties do not simultaneously show strong levels of correlation with galaxy properties and halo clustering.
For example, halo spin, which is also strongly correlated with halo clustering \citep[e.g.,][]{Mao2018}, was recently found to be ineffective in predicting galaxy spin and size \citep{jiang2019}, contrary to previous belief.
Our foremost objective being to determine whether any form of galaxy assembly bias exists, concentration is a favourable choice.
It may be interesting to investigate galaxy assembly bias signals with respect to other halo properties, that are intrinsically weaker, when larger data volumes become available.

There are several galaxy–halo connection models, besides the decorated HOD, that incorporate adjustable galaxy assembly bias, e.g., the interpolated abundance matching model in \citet{Lehmann2017}, the extended HOD model in \citet{salcedo2020}, etc.
The advantage of the decorated HOD is that it treats central and satellite galaxies separately and the values of the assembly bias parameters are straightforward to interpret.
We have shown that a mass-only model such as the HOD, though relatively flexible and successful in reproducing two-point clustering behaviour, struggles to explain our combined statistics in certain galaxy samples.
We therefore tentatively expect that signals of galaxy assembly bias would also be evident in constraints on alternative models, and we leave a detailed investigation to future work.

\subsection{Comparison to Other Work}

In relation to our primary results regarding assembly bias, there are few previous observational works with which to compare.
\citet{zentner_etal19} and \citet{vakili_2019} studied two-point clustering constraints on assembly bias using a forward-modelling approach.
Our results are in broad general agreement with these previous studies.
Both \citet{zentner_etal19} and \citet{vakili_2019} observed marginal indications of central galaxy assembly bias in the $M_r<-19.5$, $M_r<-20$, and $M_r<-20.5$ threshold samples of the same sense as our detections.
Our constraints on $\Acen$ and $\Asat$ are typically substantially more restrictive than the constraints of \citet{zentner_etal19} or \citet{vakili_2019}.
Taken at face value, our results indicate that $\Acen>0$ (positive assembly bias) for the $M_r<-20$ sample with at least $4.4\sigma$ significance. 

A number of additional authors have studied the clustering of the SDSS DR7 sample using an HOD approach, without studying assembly bias in those samples.
These include \citet{zehavi_etal11}, \citet{Guo2015wp} (who studied redshift-space clustering), and \citet{szewciw2021} (who studied projected clustering, redshift-space clustering, group multiplicity, and several other observables as part of a campaign to constrain the standard HOD optimally).
Our constraints on the standard HOD parameters are broadly consistent with the constraints from all of these studies. 
The precision with which we constrain the HOD parameters is generally better than that from \citet{zehavi_etal11} and \citet{Guo2015wp} and lies between the ``NWG'' and ``OPT'' analyses in \citet{szewciw2021}, with the ``OPT'' constraints often being a factor of 2-3 more restrictive than ours.
As we were completing this work, we became
aware of an independent study that analyses SDSS DR7 samples, building upon the framework of \citet{szewciw2021},
which reaches broadly similar conclusions \citep{beltz-mohrmann_etal22}.
In particular, they also find evidence for positive central assembly bias in the samples they study.

While our clustering only, ``$\ngal$+$\wprp$'' analyses are broadly consistent with the result of previous studies of the HOD, the introduction of cylinder counts does result in interesting deviations.
To be specific, including $\Pncic$ in the analysis drives the preferred value of $\alpha$ lower, often substantially lower than unity and substantially lower than the values quoted in previous work.
This trend is evident in our own work in \autoref{tab:fit_values}, where the ``$\ngal$+$\wprp$+$\Pncic$'' analyses yield inferred values of $\alpha$ substantially lower than the clustering-only analyses.
We suspect that this is related to the tension introduced by cluster counts and discussed above, but we relegate a full study of this shift to follow-up work.

\section{Conclusions}
\label{sec:conclusion}

We have reanalysed SDSS DR7 galaxy clustering data using a novel combination of observables and with the aim of detecting and/or constraining galaxy assembly bias. 
We used the observable combination of galaxy number density, projected two-point clustering, and counts-in-cylinders, and provided new measurements of each of these statistics.
We interpreted these data by fitting them with HOD and dHOD models of the galaxy--halo connection.
The HOD formalism is a standard "halo mass-only" model which assumes that galaxies occupy haloes in a manner that depends only upon halo mass.
Our particular implementation of the dHOD included galaxy assembly bias by allowing central and satellite occupations to vary with host halo concentration at a fixed halo mass. 
We obtained updated constraints on both the HOD and dHOD models.
In particular, we constrained the level of galaxy assembly bias, as parametrised in the dHOD model,  more precisely than any previous studies. 

Our primary conclusions are as follows.
\begin{enumerate}
    \item By adding the counts-in-cylinders statistic to the canonical analysis of number density and projected two-point function, we achieve a considerable, general improvement in the constraints on the galaxy–halo connection.\\
    
    \item We detect strong positive central galaxy assembly bias in the $M_r<-20.0$ and $M_r<-19.5$ samples, and relatively strong levels of positive central assembly bias in the $M_r<-20.5$ and $M_r<-19.0$ samples, but find no evidence of central assembly bias in the $M_r<-21.0$ sample.\\
    
    \item Negative satellite assembly bias is suggested in the $M_r<-20.0$ and $M_r<-19.0$ samples, but we do not detect satellite assembly bias in the other samples.\\
    
    \item The decorated HOD, which incorporates galaxy assembly bias, in most cases provides a better description of galaxy two-point clustering and counts-in-cylinders simultaneously than the standard mass-only HOD.
\end{enumerate}

The implications of our findings are twofold.
First, we have demonstrated the constraining power of galaxy count statistics.
There are two possible avenues for acquiring additional observational information, in order to resolve pending theoretical issues: either to use larger data sets and reduce error, or to exploit existing data using more efficient statistical probes.
The former, though very promising, especially given the great effort that is going into a new generation of galaxy surveys, does not solve the problem of degeneracies in the model parameter space, for they are intrinsic to the selected statistics, and determined by the manner of their dependence on the model.
The alternative approach, which is to develop new statistics, can be more effective in this respect.
In our case, count statistics incorporate the higher-order information of the field, which breaks multiple degeneracies with the two-point function, leading to tighter constraints in all dimensions.
We advocate the use of these statistics in combination with the commonly used two-point measurements, which has the potential not only to improve constraints on the galaxy–halo connection, but also on cosmological models and other aspects of galaxy physics.

Second, based on the dHOD formalism, we have detected galaxy assembly bias in several galaxy samples, which indicates that the first-order assumption that galaxy occupation only depends on halo mass is incomplete.
As was shown in \citet{wu08,zentner_etal14} and \citet{McCarthy_2018}, neglecting galaxy assembly bias when it is present in the real Universe may lead to systematic errors in cosmological inferences.
Our detection of galaxy assembly bias further necessitates explorations of its physical origin and better models that incorporate the effect.
Traditional models that adopt the mass-only ansatz will not suffice in the new era of precision cosmology, and developing new models that account for galaxy assembly bias is a timely task, which is attracting increasing attention \citep[see, for example,][]{yuan2021,contreras2021}.
In addition, the nature of galaxy assembly bias can be studied from several perspectives, including its dependence on cosmic time, its roots in halo assembly bias, the galaxy and halo properties that best capture the correlation, and the relevant physical processes in galaxy formation and evolution \citep[see, e.g.,][]{lacerna14,chaves-montero16,Artale_2018,contreras2019,montero-dorta2020}.
We aim to explore other applications of galaxy count statistics, and further look into the physical origin of galaxy assembly bias in future work.

\section*{Acknowledgements}
The authors thank Dhayaa Anbajagane, Camille Avestruz, Michael Blanton, Phillip Mansfield, Li-Cheng Tsai, Sihan Yuan, and Rongpu Zhou for useful discussions.

KW acknowledges support from the Leinweber Postdoctoral Research Fellowship at the University of Michigan, DoE Award DE-SC009193, and the Mellon Fellowship at the University of Pittsburgh.
YYM was supported by NASA through the NASA Hubble Fellowship grant no.\ HST-HF2-51441.001 awarded by the Space Telescope Science Institute, which is operated by the Association of Universities for Research in Astronomy, Incorporated, under NASA contract NAS5-26555. KW and ARZ were partially supported by a grant from the US National Science Foundation (NSF) through grant AST 1517563. KW, LM, and ARZ were partially supported by the Pittsburgh Particle Physics Astrophysics and Cosmology Center (PITT PACC) and the Department of Physics and Astronomy at the University of Pittsburgh. HG is supported by National Science Foundation of China (Nos. 11922305, 11833005, 12011530159). FvdB is supported by the National Aeronautics and Space Administration through Grant No. 19-ATP19-0059 issued as part of the Astrophysics Theory Program.

This research made use of Python, along with many community-developed or maintained software packages, including
IPython \citep{ipython},
Jupyter (\https{jupyter.org}),
Matplotlib \citep{matplotlib},
NumPy \citep{numpy},
SciPy \citep{scipy},
Astropy \citep{astropy},
Pandas \citep{pandas},
Corrfunc \citep{corrfunc}
and Mangle \citep{hamilton_tegmark2004, swanson2008}\footnote{We use an implementation in Python from \url{https://github.com/esheldon/pymangle}.}.
This research made use of NASA's Astrophysics Data System for bibliographic information.

\section*{Data Availability}

The data underlying this article were accessed from \url{http://sdss.physics.nyu.edu} and \url{https://www.cosmosim.org}.
The derived data generated in this research are available in the article, and any additional data will be shared on reasonable request to the corresponding author.



\bibliographystyle{mnras}
\bibliography{main} 

\begin{appendices}

\counterwithin{figure}{section}
\counterwithin{table}{section}

\section{Validation of Statistic Estimation}
\label{sec:validation}

\begin{figure*}
    \centering
    \label{fig:ngcomp}
    \includegraphics[scale=0.65]{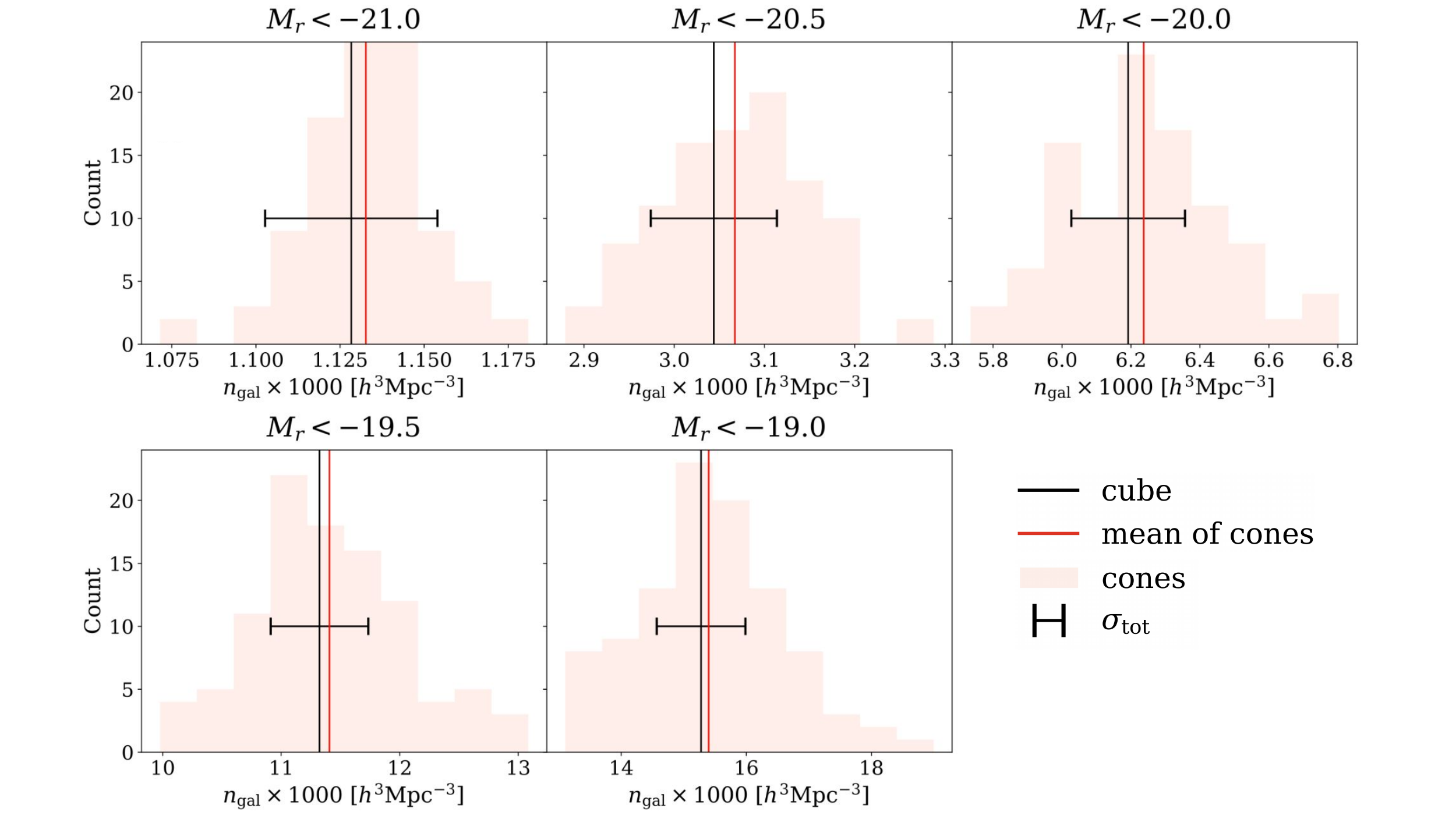}
    \caption{\textbf{Comparison between the cube mock and the cone mocks of the galaxy number density $\ngal$.}
    Each panel compares a different luminosity sample.
    The underlying cube value is shown as a vertical black line.
    The horizontal black error bars are the total jackknife error from the cube and the cones.
    The values measured from the different light cone mocks are plotted as a histogram, and the mean is marked by the vertical red line.
    This figure shows that the measurement of the galaxy number density is consistent between our cube and cone mocks for all five samples.}
\end{figure*}

\begin{figure*}
    \centering
    \label{fig:wpcomp}
    \includegraphics[scale=0.65]{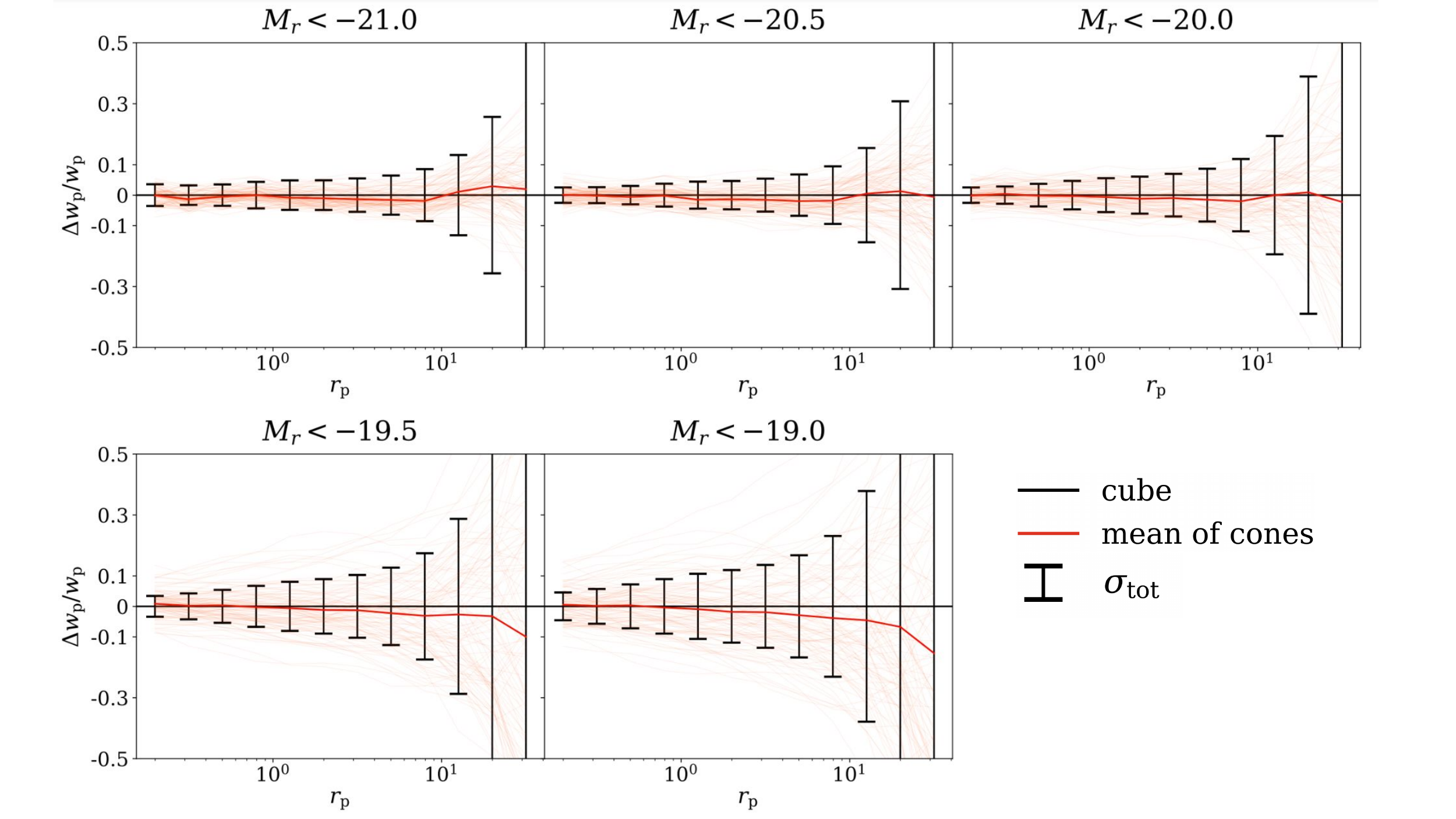}
    \caption{\textbf{Comparison between the cube mock and the cone mocks of the projected two-point function $\wp$.}
    The five panels show the fractional deviation between the cube and the cones for the different luminosity samples, labelled at the top.
    The horizontal black line marks zero deviation, and the error bars show the total jackknife error from the cube and the cones.
    Individual cones are plotted as thin lines in the background, and the mean of the cones is shown by the solid red line.
    Measurements for all the samples are consistent within error.}
\end{figure*}

\begin{figure*}
    \centering
    \label{fig:ciccomp}
    \includegraphics[scale=0.6]{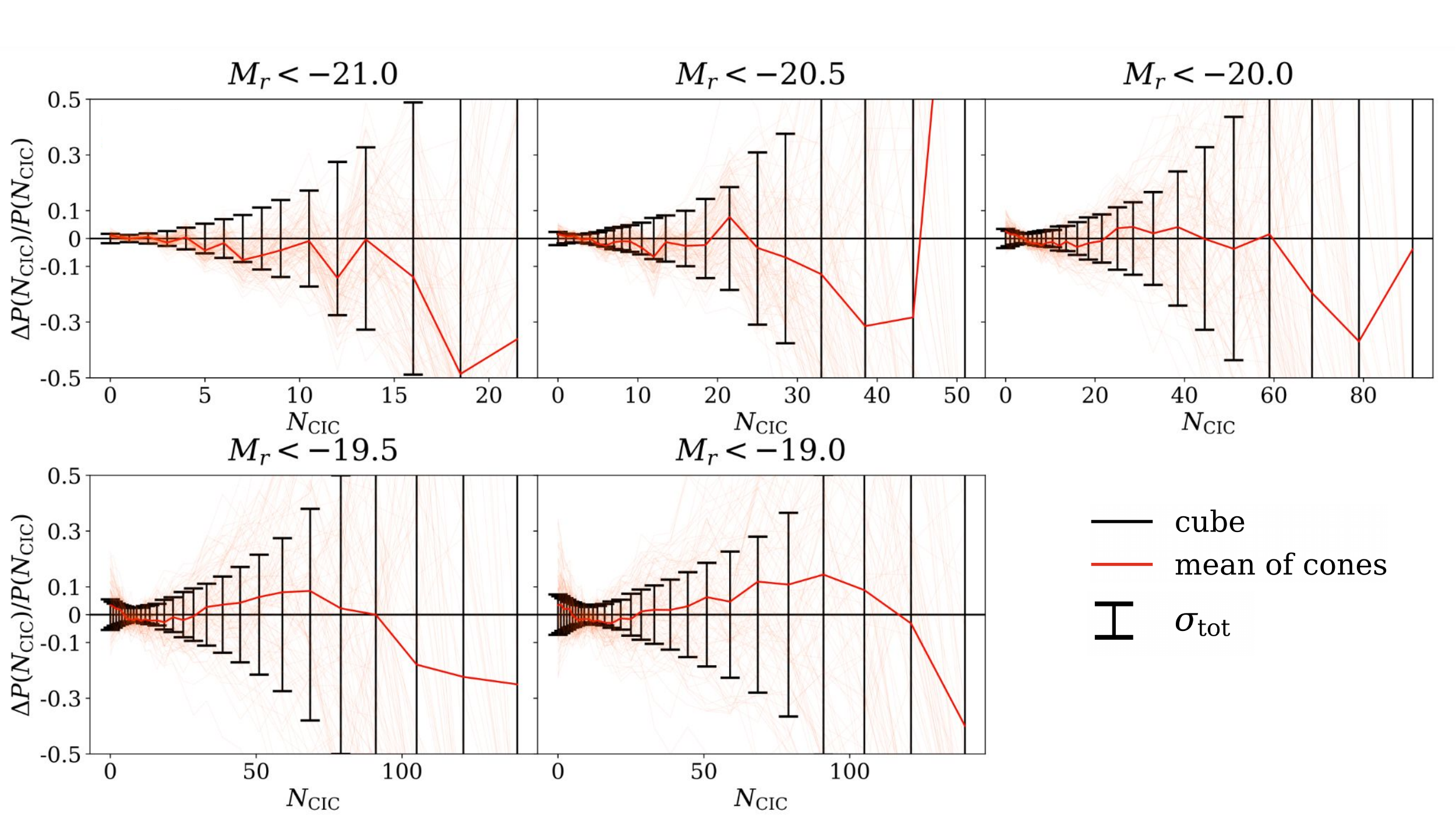}
    \caption{\textbf{Comparison between the cube mock and the cone mocks of the counts-in-cylinders statistic $\Pncic$.}
    Same as \autoref{fig:wpcomp}, but for counts-in-cylinders instead of $\wp$.
    Note that the range of the $x$-axis is different for each sample, though $y$-axis ranges are the same for all.
    Measurements are again consistent for all the samples.}
\end{figure*}

In order to fit SDSS data using a simulation, we need measurements of the observable statistics from the simulation to reflect the behaviour of real data in an unbiased manner.
However, light cone mocks that mimic observed data are computationally expensive, and therefore infeasible for sampling from the high dimensional model parameter space, whereas measurements from the original cubic volume of the simulation can be obtained much more rapidly.
Consistency between statistic measurements from the cube and the light cone mocks that have the same underlying physics would validate the use of the cube for fitting data.

In this appendix, we detail the construction of cone mocks that incorporate the relevant geometrical and observational effects present in SDSS data, obtain measurements on the cone mocks with the exact same algorithm used for SDSS data, and demonstrate that our treatment of observational effects ensures that cube mocks yield unbiased estimates of the statistics measured in cone mocks, and hence observational data.

\subsection{Halo Populating in Simulation Cube}
\label{sec:sham}
We populate the halo catalogue at $z=0.1$ from SMDPL using the stellar mass–(sub)halo mass relation model described in \citet{behroozi10} to get a galaxy catalogue in the cubic simulation volume (hereafter the ``\textbf{cube mock}'').
This model is implemented in the \texttt{halotools} software package \citep{halotools}.
The model provides the position, velocity, and stellar mass information of the galaxies.
We assume a constant mass-to-light ratio for all galaxies, such that the ranking of galaxy stellar masses from large to small is equivalent to the ranking of absolute magnitudes from bright to faint.

Since we are only interested in studying the consistency of galaxy statistics between differently constructed mocks, the specific choice of any reasonable galaxy–halo connection model in this step should not affect our results.

\subsection{Cone Building}
\label{sec:lightcone}
We build cone mocks (hereafter ``\textbf{cone mocks}'') from the cube mock described in \aref{sec:sham}, with a routine that is similar to \citet{Lange.etal.19b}.
Note that we do not populate the halo catalogue again, but directly use the cubic galaxy catalogue, such that the cube mock and cone mocks have identical underlying cosmologies and galaxy–halo connections.
The steps are as follows:
\begin{itemize}
    \item Choose a random position in the cubic volume to place our virtual observer;
    \item Periodically repeat the cube mock out to the desired depth;
    \item Calculate and record the true redshifts $\ztrue$ of the galaxies, accounting for both distance and velocity information;
    \item Apply redshift measurement uncertainty according to the model described in Appendix A of \citet{Guo2015wp}, and record the resulting redshifts with error, $\zerr$;
    \item Choose a random direction of observation, convert the galaxy positions into angular coordinates $\rm{(ra, dec)}$, and apply the 7461 $\rm{deg}^2$-SDSS footprint;
    \item Again assuming a constant mass-to-light ratio for all galaxies, rank the $r$-band apparent magnitudes of galaxies
    \item Keep galaxies with $0.01<\zerr<0.18$ and make a flux limit selection based on the apparent magnitude ranking and the total number of galaxies in the same redshift range in SDSS;
    \item Assign fibre collision status to galaxies, using the method described in \citet{guo_zehavi_zheng12}, which takes into account the tiling scheme of SDSS, and includes random collisions with background galaxies and other failures, in order to match the observed fractions of fibre collided galaxies;
    \item Perform the nearest-neighbour correction for the fibre collision effect, as was done in real data, and record the resulting $\zobs$.
\end{itemize}
We repeat this process with different random seeds to generate 100 cone mocks.

In summary, the main distinctions between the cube and cone mocks include: (i) the cone mocks mimic the angular coordinate system of observational data, whereas the cube mock adopts the plane-parallel approximation; (ii) the cone mocks account for the observational flux limit; (iii) the cone mocks have the geometry of the SDSS footprint; and (iv) the cone mocks incorporate the fibre collision effect.

\subsection{Sample Selection}
\label{sec:mockselect}
For validation purposes, we define samples in the mock catalogues by galaxy number density, that correspond to the five data samples in the first part of \autoref{tab:sample}.
This ensures similar discreteness noise in the mocks as in the real data for each sample.

For the cube mock, we multiply the number densities listed in \autoref{tab:datanwc} by the simulation volume to get the total number of galaxies $N_{\rm cube}$ in each sample.
We then select the $N_{\rm cube}$ brightest galaxies from the cube mock according to the magnitude ranking, which we identify as the luminosity-threshold samples. 
For the cone mocks, in addition to the same magnitude thresholds as in the cube, we apply the same redshift range limits on $\zobs$ as those used for selecting data samples\footnote{The galaxy luminosity function in our mock slightly differs from that in the survey, which propagates to the effective flux limit in our galaxy selection. To avoid exceeding the mock flux limit, we adopt a conservative redshift cut for the $M_r<-19.5$ threshold, at $cz_{\rm max}=23450\ \kms$ in the mocks instead of the $25450\ \kms$ limit for data, which does not affect our results qualitatively.}.

\subsection{Cube Mock Algorithm}
\label{sec:cube_algorithm}

\subsubsection{Statistics}
\label{sec:cube_stat}
In cube mocks, we apply periodic conditions and adopt the plane-parallel approximation.
We place the line of sight along the three axes in turn, apply redshift space distortions from galaxy peculiar velocities for each respectively, and average the resulting measurements for each observable.

By construction, the number density $\ngal$ of each mock sample is simply that of the corresponding data sample.
For cube measurements of $\wprp$, we use the natural estimator for $\xi(\rp, \pi)$,
\begin{equation}
\hat{\xi}_{\rm N} = \frac{\rm{DD}}{\rm{RR}} - 1,
\label{eq:xi_natural}
\end{equation}
where DD is the normalised galaxy–galaxy pair count, and RR is the normalised random–random pair count.
Given the simplicity of the geometry, we use analytic randoms instead of actually drawing random points to reduce the computational cost.
Galaxy pairs are counted in $\rp$ bins and $\xi(\rp, \pi)$ is integrated along the chosen axis out to $\pimax$.

To measure $\Pncic$, we centre a cylinder on every galaxy in the sample, and count the number of companion galaxies that fall in the cylinder, excluding the cylinder centre itself.
The histogram of the counts is calculated with our specified bins.

\subsubsection{Jackknife Covariance}
\label{sec:cube_cov}
To test the consistency between cube and cone measurements, we need to understand the uncertainty of both.
Because the cube and cone mocks have the same galaxy population, the only component of the covariance is the jackknife covariance, which provides an estimate of the fluctuation among different parts of the simulation volume.
The division of the cube into jackknife cells is trivial.
We divide the simulation volume into 100 cuboids of $40\times40\times400(\Mpch)^3$, where the long axis is the same length as the simulation, and lies along the line of sight.
We repeat the process for each of the three projections and take the average jackknife covariance.
The total covariance matrix that we use for the consistency test is the sum of the cube jackknife covariance and the cone jackknife covariance, where the cone jackknife is estimated using the scaling with volume, $\mathbf{C}_{\rm cone} = (V_{\rm cube}/V_{\rm cone})\mathbf{C}_{\rm cube}$ \citep[see, for example, Appendix A of][]{zheng_guo2016_tabcorr}.

\subsection{Comparison Between Cube and Cone Mocks}
\label{sec:compare}

We compare the measurements of $\ngal$ (\autoref{fig:ngcomp}), $\wprp$ (\autoref{fig:wpcomp}), and $\Pncic$ (\autoref{fig:ciccomp}) between the cube mock and the mean of the cone mocks generated from it.
We consider the measurements consistent between the cube and the cones if the deviation is within the total jackknife error.
In the figures, we show the cube measurements with error bars, individual cone measurements, and their mean.
We find that for all the luminosity samples we consider, and for all three of our statistics, the measurements are consistent within error.
This confirms that with our algorithm, measurements from the simulation cube can be used as unbiased estimates of statistics measured from SDSS-like datasets.
In particular, we note that the nearest neighbour correction is sufficient to account for fibre collision, for the statistics that we consider.

\section{Results for Alternative Galaxy Samples}
\label{sec:alt_samples}

\begin{table*}[]
\caption{\textbf{Inferred constraints on HOD and dHOD parameters for alternative galaxy samples.}
Same as \autoref{tab:fit_values}, but for the alternative samples that exclude the Sloan Great Wall (see \autoref{tab:sample}).}
\label{tab:fit_values_alt}
\begin{tabular}{llllllll}
\hline\hline
$M_r<-20.0^*$                                                       & $\log\Mmin$                & $\slogM$                  & $\alpha$                  & $\log\Mone$                & $\log\Mzero$               & $\Acen$                   & $\Asat$                    \\ \hline
\begin{tabular}[c]{@{}l@{}}$\ngal+\wp$\\ HOD\end{tabular}         & $12.081^{+0.323}_{-0.110}$ & $0.487^{+0.440}_{-0.315}$ & $1.099^{+0.048}_{-0.061}$ & $13.295^{+0.038}_{-0.056}$ & $10.744^{+1.172}_{-1.184}$ & -- --                     & -- --                         \\
\begin{tabular}[c]{@{}l@{}}$\ngal+\wp+\Pncic$\\ HOD\end{tabular}  & $12.164^{+0.076}_{-0.105}$ & $0.578^{+0.098}_{-0.128}$ & $1.034^{+0.032}_{-0.050}$ & $13.294^{+0.033}_{-0.064}$ & $11.084^{+1.188}_{-1.425}$ & -- --                     & -- --                         \\
\begin{tabular}[c]{@{}l@{}}$\ngal+\wp$\\ dHOD\end{tabular}        & $12.150^{+0.430}_{-0.171}$ & $0.609^{+0.502}_{-0.395}$ & $1.061^{+0.060}_{-0.077}$ & $13.291^{+0.052}_{-0.077}$ & $10.846^{+1.147}_{-1.246}$ & $0.549^{+0.330}_{-0.652}$ & $-0.060^{+0.472}_{-0.436}$ \\
\begin{tabular}[c]{@{}l@{}}$\ngal+\wp+\Pncic$\\ dHOD\end{tabular} & $12.105^{+0.110}_{-0.107}$ & $0.511^{+0.136}_{-0.166}$ & $0.955^{+0.076}_{-0.081}$ & $13.288^{+0.051}_{-0.096}$ & $11.899^{+0.592}_{-1.773}$ & $0.492^{+0.329}_{-0.441}$ & $-0.080^{+0.621}_{-0.740}$ \\ \hline
$M_r<-19.5^*$                                                       & $\log\Mmin$                & $\slogM$                  & $\alpha$                  & $\log\Mone$                & $\log\Mzero$               & $\Acen$                   & $\Asat$                    \\ \hline
\begin{tabular}[c]{@{}l@{}}$\ngal+\wp$\\ HOD\end{tabular}         & $11.918^{+0.417}_{-0.175}$ & $0.623^{+0.519}_{-0.409}$ & $1.072^{+0.046}_{-0.059}$ & $13.047^{+0.045}_{-0.066}$ & $10.821^{+1.170}_{-1.240}$ & -- --                     & -- --                      \\
\begin{tabular}[c]{@{}l@{}}$\ngal+\wp+\Pncic$\\ HOD\end{tabular}  & $11.731^{+0.042}_{-0.031}$ & $0.295^{+0.117}_{-0.136}$ & $0.868^{+0.076}_{-0.077}$ & $12.784^{+0.115}_{-0.128}$ & $12.654^{+0.139}_{-0.196}$ & -- --                     & -- --                      \\
\begin{tabular}[c]{@{}l@{}}$\ngal+\wp$\\ dHOD\end{tabular}        & $11.981^{+0.431}_{-0.233}$ & $0.723^{+0.512}_{-0.474}$ & $1.041^{+0.057}_{-0.075}$ & $13.043^{+0.063}_{-0.086}$ & $10.918^{+1.146}_{-1.301}$ & $0.506^{+0.352}_{-0.626}$ & $-0.066^{+0.492}_{-0.445}$ \\
\begin{tabular}[c]{@{}l@{}}$\ngal+\wp+\Pncic$\\ dHOD\end{tabular} & $11.717^{+0.034}_{-0.024}$ & $0.254^{+0.114}_{-0.136}$ & $0.808^{+0.071}_{-0.075}$ & $12.760^{+0.107}_{-0.124}$ & $12.671^{+0.143}_{-0.167}$ & $0.307^{+0.430}_{-0.538}$ & $-0.237^{+0.574}_{-0.541}$ \\ \hline\hline
\end{tabular}
\end{table*}

\begin{table*}[]
\caption{\textbf{Model selection and information criteria for alternative galaxy samples.}
Same as \autoref{tab:fit_stats}, but for the alternative samples that exclude the Sloan Great Wall.}
\begin{tabular}{l|lllllllll}
\hline\hline
                  & \multicolumn{3}{l}{\hspace{23pt}$\ngal+\wprp$} & & \multicolumn{3}{l}{\hspace{1pt}$\ngal+\wprp+\Pncic$} && \multirow{2}{*}{Tension} \\
                  & $\chi^2$/DoF   & AIC     & BIC    && $\chi^2$/DoF     & AIC       & BIC       &&                          \\ \hline
$M_r<-20.0^*$, HOD  & 0.99                & 17.92      & 20.75      && 1.26                & 50.21      & 58.27      && 0.31$\sigma$                 \\
\hspace{43.5pt} dHOD & 1.09                & 20.55      & 24.51      && 1.32                & 53.45      & 64.73      && 0.03$\sigma$                 \\ \hline
$M_r<-19.5^*$, HOD  & 1.23                & 19.86      & 22.68      && 1.13                & 50.56      & 59.13      && 1.33$\sigma$                 \\
\hspace{43.5pt} dHOD & 1.50                & 23.03      & 26.98      && 1.17                & 53.77      & 65.77      && 1.25$\sigma$                 \\ \hline\hline
\end{tabular}
\label{tab:fit_stats_alt}
\end{table*}

In this appendix, we list the fitting results for the alternatively selected galaxy samples for the $M_r<-20.0$ and $M_r<-19.5$ luminosity thresholds, which are described in \autoref{tab:sample}.
The alternative samples exclude the Sloan Great Wall by adopting lower depth limits, and therefore have smaller volumes and less constraining power.

The inferred HOD and dHOD parameters are listed in \autoref{tab:fit_values_alt}, with the same conventions as in \autoref{tab:fit_values}.
The model performance criteria are presented in \autoref{tab:fit_stats_alt}, which is similar to \autoref{tab:fit_stats}.

For the alternative samples, we do not find significance evidence for non-zero galaxy assembly bias.
The only case where a positive $\Acen$ is marginally suggested is the ``$\ngal$+$\wprp$+$\Pncic$'' analysis for the alternative $M_R<-20.0$ sample.
We obtain weaker constraints on the parameter space in general, relative to the constraints from the main samples.
It is unclear to what extent these differences are due to the decrease in sample size or the exclusion of the Sloan Great Wall.

\end{appendices}

\label{lastpage}
\end{document}